\documentclass[twocolumn]{aastex63}

\submitjournal{the \textit{Astrophysical Journal Supplement Series} on January 13, 2020}

\shorttitle{Methane for HITEMP}
\shortauthors{Hargreaves et al.}

\begin{document}

\title{An accurate, extensive, and practical line list of methane for the HITEMP database}

\correspondingauthor{Robert J. Hargreaves}
\email{robert.hargreaves@cfa.harvard.edu}

\author{Robert J. Hargreaves}
\affiliation{Center for Astrophysics \textbar\ Harvard \& Smithsonian,  Atomic and Molecular Physics Division, 60 Garden Street, Cambridge, MA 02138, USA}

\author{Iouli E. Gordon}
\affiliation{Center for Astrophysics \textbar\ Harvard \& Smithsonian,  Atomic and Molecular Physics Division, 60 Garden Street, Cambridge, MA 02138, USA}

\author{Michael Rey}
\affiliation{Groupe de Spectrom\'{e}trie Mol\'{e}culaire et Atmosph\'{e}rique, UMR CNRS 7331, BP 1039, F-51687, Reims Cedex 2, France}

\author{Andrei V. Nikitin}
\affiliation{V.E. Zuev Institute of Atmospheric Optics, Laboratory of Theoretical Spectroscopy, Russian Academy of Sciences, 1 Akademichesky Avenue, 634055 Tomsk, Russia}

\author{Vladimir G. Tyuterev}
\affiliation{Groupe de Spectrom\'{e}trie Mol\'{e}culaire et Atmosph\'{e}rique, UMR CNRS 7331, BP 1039, F-51687, Reims Cedex 2, France}
\affiliation{QUAMER laboratory, Tomsk State University, 36 Lenin Avenue, 634050 Tomsk, Russia}

\author{Roman V. Kochanov}
\affiliation{V.E. Zuev Institute of Atmospheric Optics, Laboratory of Theoretical Spectroscopy, Russian Academy of Sciences, 1 Akademichesky Avenue, 634055 Tomsk, Russia}
\affiliation{QUAMER laboratory, Tomsk State University, 36 Lenin Avenue, 634050 Tomsk, Russia}

\author{Laurence S. Rothman}
\affiliation{Center for Astrophysics \textbar\ Harvard \& Smithsonian,  Atomic and Molecular Physics Division, 60 Garden Street, Cambridge, MA 02138, USA}

\begin{abstract}

A methane line list for the HITEMP spectroscopic database, covering 0-13,400$\:$cm$^{-1}$ ($>$746$\:$nm), is presented. To create this compilation, \textit{ab initio} line lists of $^{12}$CH$_{4}$ from Rey et al. (2017) ApJ, 847, 105 (provided at separate temperatures in the TheoReTS information system), are now combined with HITRAN2016 methane data to produce a single line list suitable for high-temperature line-by-line calculations up to 2000$\:$K.  An effective-temperature interpolation model was created in order to represent continuum-like features at any temperature of interest. This model is advantageous to previously-used approaches that employ so-called ``super-lines'', which are suitable only at a given temperature and require separate line lists for different temperatures. The resultant HITEMP line list contains $\sim$32 million lines and is significantly more flexible than alternative line lists of methane, while accuracy required for astrophysical or combustion applications is retained. Comparisons against experimental observations of methane absorption at high temperatures have been used to demonstrate the accuracy of the new work.  The line list includes both strong lines and quasi-continuum features and is provided in the common user-friendly HITRAN/HITEMP format, making it the most practical methane line list for radiative transfer modeling at high-temperature conditions.


\end{abstract}

\keywords{brown dwarfs --- exoplanet atmospheres --- high resolution spectroscopy --- methane --- molecular spectroscopy --- radiative transfer}

\section{Introduction} \label{sec:intro}

On Earth, atmospheric methane (CH$_{4}$) is a prominent greenhouse gas that has seen a steady increase over the last decade \citep{2019Sci...364..932F}. Terrestrial CH$_{4}$ has both natural and anthropogenic sources, with atmospheric monitoring of CH$_{4}$ typically achieved using infrared spectral observations \citep{2016...acp-16-14371}. CH$_{4}$ is also the main constituent of natural gas, and plays a central role in combustion. At high temperatures, CH$_{4}$ spectra can be used for diagnostics of hydrocarbon combustion processes throughout the infrared \citep{1996ApOpt..35.4026N, 2011MeScT..22b5303P, 2015JQSRT.155...66S, 2019arXiv191008116T}. 

Beyond terrestrial environments, CH$_{4}$ has been identified in the spectra of numerous sub-stellar astrophysical environments \citep{1978Natur.273..281H, 1991ApJ...376..556L, 1996Sci...272.1310M, 2018Icar..300..174Y}. CH$_{4}$ absorption in the 1.0-2.5$\:\mu$m region is the characterizing feature of T-type brown dwarfs \citep{1995Sci...270.1478O, 2005ARA&A..43..195K,2015MNRAS.450..454C} with effective temperatures of $\sim$500-1400$\:$K \citep{2014PASA...31...43B}.  This attribute can be exploited to identify T dwarfs through `methane imaging' \citep{2018ApJS..236...28T}. For mid-to-late L dwarfs, CH$_{4}$ absorption can remain observable near 3.3$\:\mu$m for higher temperatures  \citep{2000ApJ...541L..75N, 2009ApJ...702..154S}.  As the temperature drops, CH$_{4}$ absorption remains dominant in the spectra of Y dwarfs  \citep{2011ApJ...743...50C, 2012ApJ...753..156K} and is also present in the atmospheres of the giant planets \citep{2005Icar..176..255I, 2008SSRv..139..191M} and Titan \citep{1994Icar..111..174K, 2006.ATREYA20061177}.

Since the detection of 51 Pegasi b \citep{1995Natur.378..355M}, there are now in excess of 4000 known exoplanets. Studies of transiting exoplanets have been able to probe the atmospheres of a small number of these objects \citep{2018AJ....155..156T}, with observations of water vapor \citep{2008Natur.456..767G} and carbon monoxide absorption \citep{2013Sci...339.1398K}. Models predict CH$_{4}$ to be more abundant than carbon monoxide below $\sim$1300$\:$K \citep{1999ApJ...512..843B}, yet observations of CH$_{4}$ have only been reported in the spectra of five exoplanets to date: HD 189733b \citep{2008Natur.452..329S}, HD 209458b \citep{2009ApJ...704.1616S}, XO-1b \citep{2010ApJ...712L.139T}, HR 8799b \citep{2015ApJ...804...61B} and 51 Eridani b \citep{2015Sci...350...64M}.

Many exoplanet observations have used instruments with low resolving powers \citep{2019AJ....157..114B}, where $R=\lambda/\Delta\lambda\lesssim200$, which can limit the capability to identify individual molecular species. However, recent spectroscopic techniques such as cross-correlation \citep{2014Natur.509...63S} and Doppler tomography \citep{2019MNRAS.tmp.2316W} are able to take advantage of high resolution instruments ($R\sim25,000 - 100,000$) to definitely confirm detections of H$_{2}$O \citep{2017AJ....153..138B},  CO \citep{2010Natur.465.1049S}, TiO \citep{2017AJ....154..221N}, as well as neutral and ionized atoms \citep{2018Natur.560..453H}, from exoplanet transit spectra. These methods have also highlighted the need for line lists to be both accurate and complete at high resolutions \citep{2015A&A...575A..20H}.

The pressing need for improvements to line lists for planetary spectroscopy (including  CH$_{4}$) have been emphasized in a number of review papers \citep{2013A&ARv..21...63T, 2014RSPTA.37230087B, 2016arXiv160206305F, 2017MolAs...8....1T, 2019astro2020T.146F}. These improvements are essential to make the most of measurements from the forthcoming \textit{Atmospheric Remote-sensing Infrared Exoplanet Large-survey (ARIEL)} mission \citep{2018ExA....46..135T}, which is dedicated to exoplanet observations. Furthermore, the \textit{James Webb Space Telescope} will provide a significant advancement in the capability to characterize exoplanet atmospheres using moderate resolution ($R\lesssim3500$) spectroscopy \citep{2016ApJ...817...17G}.

\subsection{Methane spectroscopy} \label{sec:intro:spec}

The polyad nature of CH$_{4}$ is a consequence of all four vibrational modes having the relationship $\nu_{1}\approx\nu_{3}\approx2\nu_{2}\approx2\nu_{4}\approx3000\:$cm$^{-1}$. Each polyad is identified by $P_{n}$, where $n=2(v_{1}+v_{3})+v_{2}+v_{4}$ (with $v_{i}$ equal to the number of quanta of each mode), but named according to the number of vibrational bands within each polyad. For example, the second polyad $P_{2}$ contains 5 vibrational bands ($\nu_{1}$, $\nu_{3}$, $2\nu_{2}$, $2\nu_{4}$, $\nu_{2}+\nu_{4}$), and is therefore referred to as the pentad \citep{2006JQSRT..98..394B}. Due to the tetrahedral symmetry of the CH$_{4}$ molecule, the degenerate overtone and combination vibration states involved in successive polyads are split into sub-levels, which complicates ro-vibrational band patterns for analyses. Early versions of spectroscopic databases specifically developed for CH$_{4}$ and other high-symmetry molecules, such as TDS \citep{1994JQSRT.TDS}, STDS \citep{1998JQSRT.STDS} and MeCaSDa \citep{2013JQSRT.130...62B}, have been constructed using empirical effective models for isolated polyads.  

The HITRAN2016 database \citep{2017JQSRT.203....3G} details the most accurate collection of line parameters for CH$_{4}$, with a primary focus towards the modeling of the terrestrial atmosphere. This is also the focus of the GEISA \citep{2016JMoSp.327...31J}, MeCaSDa \citep{2013JQSRT.130...62B} and GOSAT \citep{2015JQSRT.154...63N} databases. These linelists, which are based on experimental measurements and/or empirical fits of laboratory spectra, suffer from incompleteness issues for high-temperature conditions because of insufficient information on experimentally measured and assigned transitions. They are therefore unsuitable for astrophysical applications with a large range of temperatures.

Assigning individual transitions becomes a significant challenge in dense spectra with numerous blended features, as is the case for CH$_{4}$. Since HITRAN2016, there has been steady progress in assigning room-temperature and lower-temperature spectra \citep{2017JQSRT.203..341N,  2018JQSRT.219..323N, 2019JQSRT.225..351R, 2019JQSRT.23906646N}. Many of these studies, as well as HITRAN2016 updates, have already benefited from supplementary information for the resonance interaction parameters within vibrational polyads. These are derived from an \textit{ab initio} potential energy surface that made analyses of experimental spectra more consistent and reliable, as described in \citet{2013JPC.CTmethane}. However this was only done for cold bands and for relatively low polyads up to $\sim$7300$\:$cm$^{-1}$. The difficulty of extending assignments is strongly exacerbated at higher temperatures. For this reason, a number of high-temperature laboratory measurements have been made of CH$_{4}$ in both emission \citep{2003JQSRT..82..279N, 2008JQSRT.109.2027T, 2012ApJ...757...46H, 2018JChPh.148m4306A, 2018Amyay.erratum, 2019RScI...90i3103G} and absorption \citep{2014JMoSp.303....8A, 2015ApJ...813...12H, 2018JQSRT.215...59G, 2019ApJS..240....4W}.

On the theoretical side, the hot bands and high-$J$ transitions have been included in global variational CH$_{4}$ line lists : `10to10' \citep{2014MNRAS.440.1649Y}, as part of the ExoMol project \citep{2016JMoSp.327...73T}, and the \citet{2014ApJ...789....2R} line list (referred to here as RNT2014) as a part of TheoReTS project \citep{2016JMoSp.327..138R}. These works demonstrated that  \textit{ab initio} line lists of CH$_{4}$ could approach the accuracy required for high-temperatures, but the inclusion of billions of transitions made the resulting full line-by-line lists impractical for typical applications. When comparing these line lists, \citet{2015ApJ...813...12H} recommended the separation of strong and continuum-like features. Indeed, it was shown by \citet{2014ApJ...789....2R} that it is necessary to account for approximately 1 million rovibrational transitions per 1$\:$cm$^{-1}$ for CH$_{4}$ opacity calculations at 2000$\:$K. To make online computations of the absorption cross-section faster, it was suggested to model the quasi-continuum formed by the contributions of huge amounts of very weak lines using so called ``super-lines'', as originally implemented in the TheoReTS database \citep{2016JMoSp.327..138R}.  Super-lines represent integrated intensity contributions from tiny transitions on a pre-defined grid of small wavenumber and temperature intervals.

Updated  state-of-the-art \textit{ab initio} line lists have since been published, ExoMol `34to10' \citep{2017A&A...605A..95Y} and \citet{2017ApJ...847..105R} (referred to here as RNT2017), both of them using the super-line approach for the compression of relatively weak absorption/emission features complemented with lists of medium and strong lines. To obtain the full CH$_{4}$ spectrum, both the strong and super-line components are required. In each case, these line lists still require a large quantity of strong lines to cover the temperature range of calculations. Furthermore, a separate super-line component is provided at each temperature, which makes them difficult to integrate into existing radiative transfer codes and significantly less flexible than a standard line list.

\subsection{The HITRAN and HITEMP databases} \label{sec:intro:hitemp}

The HITRAN database contains detailed spectroscopic line-by-line parameters of 49 molecules with many of their isotopologues (along with absorption cross-sections for almost 300 molecules, collision-induced absorption spectra for many collisional pairs, and aerosol properties). HITRAN2016 \citep{2017JQSRT.203....3G} is the most recent version of the database, and is freely available at HITRAN\textit{online}\footnote{https://hitran.org}. Recent efforts have been undertaken to expand the use of HITRAN towards planetary atmospheres, with the inclusion of additional broadening species \citep{2016JQSRT.168..193W, 2019...Tan..H2..HIT}. However, the CH$_{4}$ line list in HITRAN2016 is unsuitable for spectroscopy at high temperatures due to issues of incompleteness. This is a consequence of the absence of many vibrational hot bands, high ro-vibrational transitions or any other extremely weak transitions (at terrestrial temperatures), due to their negligible effect in terrestrial atmospheric applications.

The HITEMP database \citep{2010JQSRT.111.2139R} was established specifically to model gas-phase spectra in high-temperature applications, and can be thought of as a ``sister'' to HITRAN (with data also provided through HITRAN\textit{online}). One substantial difference between HITRAN and HITEMP is the number of transitions included for each molecular line list, a consequence of the inclusion of numerous vibrational hot bands, high ro-vibrational transitions and overtones. This difference is most apparent for H$_{2}$O, where there are currently $\sim$800 times the number of lines in HITEMP2010 when compared to HITRAN2016. Typically, these additional transitions constitute numerous lines (often millions) from \textit{ab initio} or semi-empirical calculations, which are then combined with accurate parameters from HITRAN. The HITEMP database has been undergoing a large scale update \citep{2015ApJS..216...15L, 2019JQSRT.232...35H} and, prior to this work, included seven molecules: H$_{2}$O, CO$_{2}$, N$_{2}$O, CO, NO, NO$_{2}$, and OH.

For HITRAN and HITEMP, the temperature-dependent spectral line intensity of a transition, $\nu_{ij}$ (cm$^{-1}$), between two rovibronic states is given as
\begin{equation}\label{eqn_line strength}
  S_{ij}(T) = \frac{A_{ij}}{8\pi c \nu_{ij}^{2}}  \frac{ g^{\prime} I_{a}}{ Q(T) } \textrm{exp}\left(\frac{-c_{2}E^{\prime\prime}}{T}\right) \left[ 1 - \textrm{exp}\left(\frac{-c_{2}\nu_{ij}}{T} \right) \right] \; ,
\end{equation}
where $A_{ij}$ (s$^{-1}$) is the Einstein coefficient for spontaneous emission, $g^{\prime}$ is the upper state statistical weight, $E^{\prime\prime}$ (cm$^{-1}$) is the lower-state energy, $Q(T)$ is the total internal partition sum, $I_{a}$ is the natural terrestrial isotopic abundance\footnote{One should note that isotopic abundance is dependent upon the environment and HITRAN is consistent with specific terrestrial values given by \citet{1984JPCRD..13..809D}. For applications that do not assume these isotopic mixtures (e.g., exoplanetry atmospheres), this weighting should be renormalized by the user.}, and $c_{2} = hc/k = 1.4387770\:$cm$\:$K, the second radiation constant. To remain consistent, the spectroscopic parameters in HITRAN and HITEMP are provided at a reference temperature of 296$\:$K and the line intensities are scaled to terrestrial abundances. The units\footnote{Line positions in HITRAN and HITEMP are provided in reciprocal centimeter (cm$^{-1}$) and denoted $\nu$ (thereby dropping the tilde that is the official designation of wavenumber, $\tilde{\nu}$), and pressure in atm (atmosphere). Intensity is traditionally expressed as cm$^{-1}$/(molecule$\:$cm$^{-2}$) rather than simplifying to the equivalent cm$\:$molecule$^{-1}$.} used throughout HITRAN editions do not strictly adhere to the SI system for both historical and application-specific reasons.

The HITRAN Application Programming Interface, HAPI \citep{2016JQSRT.177...15K}, is available via HITRAN\textit{online} and is provided for users to work with the HITRAN and HITEMP line lists. The line-by-line nature and consistency between the HITRAN and HITEMP databases mean that they are extremely flexible when modeling a variety of environments. The HITRAN and HITEMP parameters undergo rigorous validations against observations \citep{2019JQSRT.23606590O, 2019JQSRT.232...35H} and are regularly used in radiative transfer codes such as LBLRTM \citep{2005JQSRT..91..233C}, NEMESIS \citep{2008JQSRT.109.1136I}, the Reference Forward Model \citep{2017JQSRT.186..243D}, RADIS \citep{2019JQSRT.222...12P} and the Planetary Spectrum Generator \citep{2018JQSRT.217...86V}.

This article describes the addition of CH$_{4}$ to the HITEMP database, bringing the total number of HITEMP molecules to eight. The aim of this line list is to be accurate and complete, but at same time practical (in terms of time required to calculate opacities) for high-temperature applications.

\section{Line lists compared in this work} \label{sec:lists}
Over the last decade, there has been a significant increase in the capability of theoretical calculations for CH$_{4}$ spectroscopy at high temperatures \citep{2014ApJ...789....2R,2014MNRAS.440.1649Y,2017ApJ...847..105R,2018A&A...614A.131Y}, which coincides with the requirement for sufficiently accurate high-temperature line lists in order to characterize brown dwarfs and exoplanets \citep{2017MolAs...8....1T, 2019astro2020T.146F}. This article broadly describes the three state-of-the-art line lists of CH$_{4}$ that have been used (and compared) in this work.

\subsection{HITRAN2016} \label{subsec:lists:hitran}

In HITRAN2016 \citep{2017JQSRT.203....3G}, CH$_{4}$ (molecule 6) contains parameters for four isotopologues: $^{12}$CH$_{4}$, $^{13}$CH$_{4}$, $^{12}$CH$_{3}$D and $^{13}$CH$_{3}$D. Line parameters are provided at 296$\:$K and intensities are scaled for natural abundances (0.988274, 0.011103, 6.15751$\times$10$^{-4}$ and 6.91785$\times$10$^{-6}$, respectively). The partition function from \citet{2017JQSRT.203...70G} is recommended when using HITRAN2016, and is also provided at HITRAN\textit{online}.

For $^{12}$CH$_{4}$ there are 313,943 transitions up to 11,502$\:$cm$^{-1}$ ($P_{8}$). Below 6230$\:$cm$^{-1}$, there are both upper-state and lower-state assignments for vibrational and rotational quanta for almost all transitions, however there are only limited assignments beyond 6230~cm$^{-1}$. The majority of assigned transitions have been validated in laboratory experiments, with weaker features being provided from calculated line lists such as MeCaSDa \citep{2013JQSRT.130...62B}. \citet{2012Icar..219..110C} provide $\sim$2500 assignments between 6230-7920$\:$cm$^{-1}$. For unassigned lines in this region, $E^{\prime\prime}$ has been determined for approximately half of these lines from spectra at 80 and 300$\:$K, and remaining lines contain an estimated $E^{\prime\prime}$. Between 7920-10,450$\:$cm$^{-1}$, empirical line positions and intensities are provided without assignments and with a constant $E^{\prime\prime}$ \citep{2005JQSRT..96..251B, 2015JMoSp.308....1B, 2015JQSRT.166....6B}. Finally, limited lower rotational assignments are given for lines between 10,920-11,502$\:$cm$^{-1}$ \citep{2013...Priv..Benner}.

For all spectral ranges, line-shape parameters have been provided from appropriate empirical observations. When these were unavailable, line-shape parameters have been calculated using the algorithms described by \citet{2013JQSRT.130..201B} and \citet{2009JQSRT.110..654L}.

The main issue for the modeling of CH$_{4}$ absorption/emission at elevated temperature is to account for the rapidly increasing contributions of hot bands, in which a huge amount of excited rovibrational levels for high-energy polyads \citep{2013JPC.CTmethane, 2015JQSRT.PFmethane, 2017ApJ...847..105R} are involved. As mentioned, HITRAN2016 is unsuitable for high-temperature applications due to lack of completeness for hot bands and high-$J$ transitions, but also because the assignment deficiencies and limited knowledge of lower-state energies, $E^{\prime\prime}$, for large spectral regions introduce errors at temperatures beyond room-temperature. This is particularly true for the portion of the line list beyond 6230$\:$cm$^{-1}$ (i.e., $<1.3\:\mu$m).

\subsection{RNT2017 and TheoReTS calculated data} \label{subsec:lists:theorets}

For this study we use RNT2017, the latest high-temperature theoretical line list for $^{12}$CH$_{4}$ constructed by \citet{2017ApJ...847..105R} and provided as part of the Reims-Tomsk collaboration via the TheoReTS data system \citep{2016JMoSp.327..138R}. RNT2017 contains significant improvements with respect to the previous RNT2014 \citep{2014ApJ...789....2R}  work, for which a good general agreement with experimental spectra up to 1200$\:$K has been observed by \citet{2015ApJ...813...12H} for the pentad ($P_{2}$) and octad ($P_{3}$) regions (2.0-3.8$\:\mu$m) . RNT2017 has recently been validated against experimental observations up to 1000~K for the tetradecad ($P_{4}$), icosad ($P_{5}$) and triacontad ($P_{6}$) regions (1.11-1.85$\:\mu$m) by \citet{2019ApJS..240....4W} at resolutions of 0.02, 0.2 and 2.0$\:$cm$^{-1}$. In addition, the region near 1.7~$\mu$m has also been validated to accurate ($\pm$0.002$\:$cm$^{-1}$) observations at 1000$\:$K by \citet{2018JQSRT.215...59G} along with comparisons to MeCaSDa, HITRAN2016 and ExoMol 10to10. 

The RNT2017 line list was created in three steps. The first was to provide over 150 billion transitions (with a lower-state rovibrational energy cutoff of 33,000$\:$cm$^{-1}$) from first-principles quantum mechanical variational calculations using the molecular potential energy surface of \citet{2011CPL...501..179N, 2016JChPh.145k4309N}. The line intensities were calculated from the purely \textit{ab initio} dipole moment surfaces of \citet{2017JQSRT.200...90N}. The resulting line list ranges from 0-13,400$\:$cm$^{-1}$ (i.e., $>746\:$nm) with a maximum temperature of 3000~K.

To improve the accuracy of the \textit{ab initio} line positions, a second step applied empirical corrections for 3.7 million of the strongest transitions. This involves $\sim$100,000 energy levels extracted from analyses of experimental laboratory room-temperature spectra. No empirical corrections were applied to line intensities, which were computed from an \textit{ab initio} dipole moment surface using a variational method. 

A third and final step follows the recommendation of \citet{2015ApJ...813...12H} to separate the empirically corrected line lists into two components: ``strong'' and ``super'' lines. To obtain the full CH$_{4}$ spectrum at each temperature, both the strong and super-line lists are required. The number of lines in each subsequent line list (at each temperature) is shown in Tab.~(\ref{tab_ch4_strong_lines}). Full details are described by \citet{2017ApJ...847..105R} with only important points explained here. 

From the billions of transitions that are computed, 
an intensity cutoff function, $I_{\textrm{\scriptsize{cut}}}(\nu,T)$, is used to exclude the weakest transitions that have a negligible contribution to the total opacity at each temperature. The cutoff function has the approximate structure of an extremely low-resolution CH$_{4}$ spectrum and is dependent on the wavenumber and temperature. 

To separate between strong and super-lines at each temperature,  a temperature-dependant scale factor ($\alpha_{\textrm{\scriptsize{strong}}}(T)$) is applied to the cutoff functions such that $I_{\textrm{\scriptsize{strong}}}(\nu,T) = \alpha_{\textrm{\scriptsize{strong}}}(T) I_{\textrm{\scriptsize{cut}}}(\nu,T)$. All transitions that have an intensity $I(\nu,T)>I_{\textrm{\scriptsize{strong}}}(\nu,T)$ are retained for the strong line lists. These strong lines are necessary for accurate simulation of sharp features in absorption/emission spectra. Transitions that have an intensity  $I_{\textrm{\scriptsize{strong}}}(\nu,T) > I(\nu,T) > I_{\textrm{\scriptsize{cut}}}(\nu,T)$ are compressed into so-called super-lines  \citep{2016JMoSp.327..138R}. These super-lines are provided on a 0.005 cm$^{-1}$ grid and account for billions of weak transitions. The compression of the full line list at each temperature reduces the number of lines necessary for line-by-line calculations and increases the efficiency of radiative transfer calculations. However, the downside of this compression means that the parameters of individual contributing transitions are not stored (e.g., $\nu$, $I$, $E^{\prime\prime}$, $J^{\prime\prime}$). It is also worth noting that the intensity of the super-lines can exceed $I_{\textrm{\scriptsize{strong}}}(\nu,T)$ for high temperatures: a consequence of the super-lines including predominantly hot bands and high rotational levels, which become increasingly populated at higher temperatures.

\begin{figure}[t!]
\centering
\includegraphics[scale=0.22, trim={0 0 0 0}, clip]{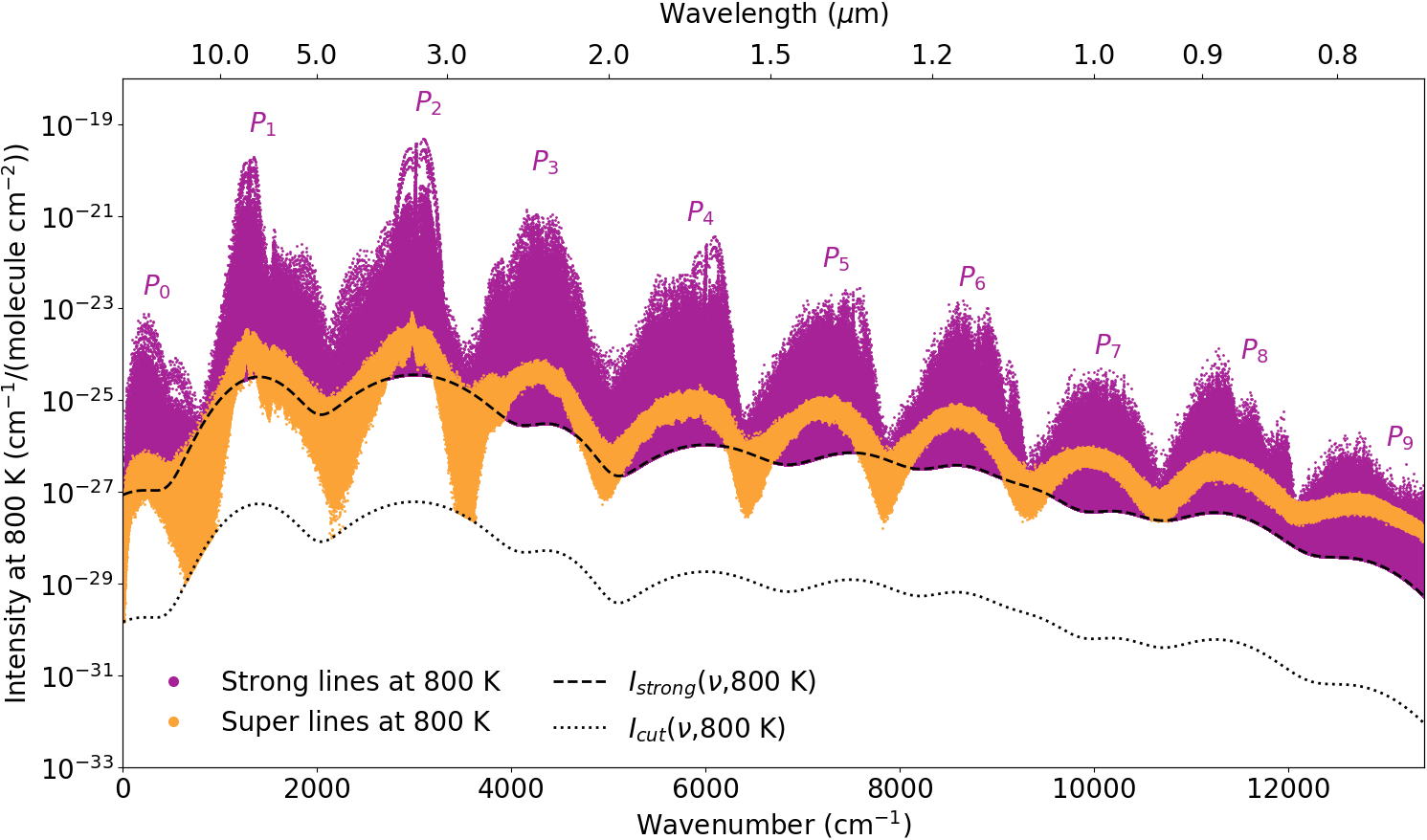} 
\caption{The intensities and positions of strong and super-lines from RNT2017 \citep{2017ApJ...847..105R} at 800$\:$K. The intensity cutoff, $I_{\textrm{\scriptsize{cut}}}$($\nu$,~800$\:$K), and strong line threshold, $I_{\textrm{\scriptsize{strong}}}$($\nu$,~800$\:$K), are given as the dashed lines. For reference, each polyad region has been indicated. \label{fig_800_strong_super}}
\end{figure}

Fig.~(\ref{fig_800_strong_super}) displays the strong and super-line components of the 800$\:$K line list, plotted alongside  $I_{\textrm{\scriptsize{strong}}}$($\nu$,~800$\:$K) and  $I_{\textrm{\scriptsize{cut}}}$($\nu$,~800~K). 
RNT2017 provides a separate strong and super-line list for each temperature, with the files used for this work summarized in Tab.~(\ref{tab_ch4_strong_lines}) along with intensity sums ($\Sigma S_{\textrm{\scriptsize{RNT}}}(T)$). A total number of 216 million lines are required for calculations between 300-2000$\:$K, of which $\sim$179 million are from the strong line lists and $\sim$37 million are from the super-line lists.

\begin{deluxetable*}{c|c|cc|cc|cc}
\tablecaption{Summary of the individual $^{12}$CH$_{4}$ line lists used in this work from \citet{2017ApJ...847..105R}. At each temperature, the number of lines ($N_{\textrm{\scriptsize{RNT}}}(T)$) and intensity sums ($\Sigma S_{\textrm{\scriptsize{RNT}}}(T)$) are given for the total line list, along with the strong and super-line components. \label{tab_ch4_strong_lines}}
\tablehead{
\colhead{}              & \colhead{} &
\colhead{}              & \colhead{$\Sigma S_{\textrm{\scriptsize{RNTstr}}}(T)$\tablenotemark{{\scriptsize b}}, } &
\colhead{}              & \colhead{$\Sigma S_{\textrm{\scriptsize{RNTsup}}}(T)$\tablenotemark{{\scriptsize b}}, } &
\colhead{}              & \colhead{$\Sigma S_{\textrm{\scriptsize{RNTtot}}}(T)$\tablenotemark{{\scriptsize b}}, } \\
\colhead{$T$}           & \colhead{$\nu_{\textrm{\scriptsize{max}}}$\tablenotemark{{\scriptsize a}}} &
\colhead{$N_{\textrm{\scriptsize{RNTstr}}}(T)$}          & \colhead{$\times 10^{-17}$} &
\colhead{$N_{\textrm{\scriptsize{RNTsup}}}(T)$}          & \colhead{$\times 10^{-18}$} &
\colhead{$N_{\textrm{\scriptsize{RNTtot}}}(T)$}          & \colhead{$\times 10^{-17}$} \\
\colhead{(K)}           & \colhead{(cm$^{-1}$)} &
\colhead{}              & \colhead{(cm$^{-1}$/(molecule$\:$cm$^{-2}$))} &
\colhead{}              & \colhead{(cm$^{-1}$/(molecule$\:$cm$^{-2}$))} &
\colhead{}              & \colhead{(cm$^{-1}$/(molecule$\:$cm$^{-2}$))}
}
\startdata
 300 &  13,400 &   1,939,483 &  1.773  &  1,734,619 &  0.008    &   3,674,102 &  1.773 \\
 400 &  13,400 &   3,064,078 &  1.774  &  2,123,246 &  0.023    &   5,187,324 &  1.776 \\
 500 &  13,400 &   3,707,529 &  1.776  &  2,401,231 &  0.054    &   6,108,760 &  1.781 \\
 600 &  13,400 &   3,801,808 &  1.776  &  2,546,247 &  0.136    &   6,348,055 &  1.790 \\
 700 &  13,400 &   5,087,143 &  1.776  &  2,645,520 &  0.239    &   7,732,663 &  1.800 \\
 800 &  13,400 &   7,452,706 &  1.775  &  2,677,728 &  0.367    &  10,130,434 &  1.812 \\
 900 &  12,600 &   6,728,693 &  1.756  &  2,519,747 &  0.662    &   9,248,440 &  1.822 \\
1000 &  12,600 &   7,638,016 &  1.730  &  2,519,825 &  1.028    &  10,157,841 &  1.833 \\
1100 &  12,000 &   9,966,742 &  1.690  &  2,399,832 &  1.537    &  12,366,574 &  1.844 \\
1200 &  11,200 &  11,701,566 &  1.637  &  2,239,890 &  2.117    &  13,941,456 &  1.849 \\
1300 &  10,700 &  13,041,320 &  1.573  &  2,139,895 &  2.842    &  15,181,215 &  1.857 \\
1400 &   9,500 &  14,784,894 &  1.502  &  1,899,906 &  3.582    &  16,684,800 &  1.860 \\
1500 &   9,500 &  14,389,334 &  1.409  &  1,899,917 &  4.500    &  16,289,251 &  1.859 \\
1600 &   8,000 &  14,591,701 &  1.323  &  1,599,953 &  5.298    &  16,191,654 &  1.853 \\
1700 &   8,000 &  14,429,314 &  1.178  &  1,599,966 &  6.589    &  16,029,280 &  1.837 \\
1800 &   8,000 &  14,511,952 &  1.050  &  1,599,969 &  7.660    &  16,111,921 &  1.816 \\
1900 &   6,600 &  15,699,493 &  0.961  &  1,319,967 &  8.239    &  17,019,460 &  1.785 \\
2000 &   6,600 &  16,051,329 &  0.861  &  1,319,972 &  9.072    &  17,371,301 &  1.768 \\
\enddata
\tablenotetext{{\scriptsize a}}{The maximum wavenumber for each line list.}
\tablenotetext{{\scriptsize b}}{Intensity sums have been scaled by 0.988274, the natural abundance of $^{12}$CH$_{4}$.}
\end{deluxetable*}

The individual RNT2017 line lists are considered complete up to the maximum wavnumber, $\nu_{\textrm{\scriptsize{max}}}$, given in Tab.~(\ref{tab_ch4_strong_lines}). Here, completeness signifies that all lines of sufficient intensity are included in the calculation. That is to say, including additional transitions has a negligible contribution to the total opacity, it is converged. For example, the RNT2017 line list at 1200$\:$K is complete up to 11,200$\:$cm$^{-1}$ with total intensity sum $\Sigma S_{\textrm{\scriptsize{RNTtot}}}$(1200$\:$K) = 1.849$\times 10^{-17}\:$cm$^{-1}$/(molecule$\:$ cm$^{-2}$). Line list extrapolation was recommended for wavenumber/temperature ranges outside of these limits by scaling the resulting super-line intensities. 
\\

\subsection{ExoMol 34to10} \label{subsec:lists:exomol}
The ExoMol project \citep{2016JMoSp.327...73T} is currently at the forefront of theoretical line list calculations for astrophysically relevant molecules, along with the NASA Ames group \citep{2017JQSRT.203..224H} and TheoReTS project (see Sect.~\ref{subsec:lists:theorets}). For  $^{12}$CH$_{4}$, the ExoMol 34to10 line list \citep{2017A&A...605A..95Y} represents an extension to the previous version, 10to10 \citep{2014MNRAS.440.1649Y}. The 10to10 line list has been compared to experimental observations of the pentad ($P_{2}$) and octad ($P_{3}$) regions up to 1200$\:$K \citep{2015ApJ...813...12H} alongside RNT2014, as well as near 1.7$\:\mu$m at 1000$\:$K  alongside RNT2017 \citep{2018JQSRT.215...59G}. In both cases, it was noted that the ExoMol line lists covered important needs for astrophysical applications,  but were not of sufficient accuracy for high-resolution applications. 

Data from the ExoMol group are regularly used to update the HITRAN and HITEMP databases \citep{2010JQSRT.111.2139R, 2017JQSRT.203....3G, 2019JQSRT.232...35H} because the \textit{ab initio} intensities for some molecules are of exceptional quality. Most notable examples include H$_{2}$O \citep{BT2_2006,Lodi2011,LodiTennyson2012} and CO$_2$ \citep{zak2016}, where ExoMol intensities are used for a large portion of the HITRAN2016 lines. While the ExoMol $^{12}$CH$_{4}$ line lists have not been included as part of this work, a brief description is provided for the reader because ExoMol 34to10 is the only other comparable line list. It is therefore used for high-temperature simulations, such as for exoplanet atmospheres \citep{2015ApJ...804...61B}, and is used for comparison here.

For the 34to10 line list, a total number of 34 billion transitions were calculated, with a maximum transition frequency of 12,000$\:$cm$^{-1}$, maximum $E^{\prime\prime}$ of 10,000$\:$cm$^{-1}$ and a temperature range up to 2000$\:$K. The line list was also partitioned into ``strong'' and ``weak'' components, with the strong lines represented by a line list of $\sim$17 million transitions and the weaker lines compressed into separate super-line lists at each temperature ($\sim$7 million per temperature). As is the case for RNT2017, to reproduce the full spectrum of CH$_{4}$ at each temperature, both  the  strong  and  super-lines  lists  are required ($\sim$71 million lines for 300-2000$\:$K).

The completeness of the 34to10 line list has improved when compared to 10to10, with the partitioning of the line list making it more practical to use. However, the underlying energy levels (and transition frequencies) have not been adjusted and therefore the accuracy issues noted for 10to10 remain relevant to 34to10. Line intensities are also significantly overestimated with respect to experimental data for high wavenumber ranges.

\section{A methane line list for HITEMP} \label{sec:list4hitemp}
HITEMP  follows the same format and  formalism as HITRAN and can therefore be easily used in existing line-by-line radiative transfer codes. A single CH$_{4}$ line list that is simultaneously accurate, extrensive and practical has been constructed by merging the combined RNT2017 and HITRAN2016 line lists.

\subsection{Combining the RNT2017 line lists} \label{sec:combining}

The first step was to combine the strong line lists from RNT2017 into a single global list. A spectral line intensity at $T_{0}$, given in Eqn.~(\ref{eqn_line strength}), can be converted to temperature $T$ using the well-known relationship
\begin{equation}\label{eqn_int_ratio}
  \frac{S_{ij}(T)}{ S_{ij}(T_{0})} =  \frac{ Q(T_{0})}{ Q(T) } \textrm{exp}\left(\frac{c_{2}E^{\prime\prime}}{T_{0}} - \frac{c_{2}E^{\prime\prime}}{T}\right) \frac{ 1 - \textrm{exp}(-c_{2}\nu_{ij}/T) }{ 1 - \textrm{exp}(-c_{2}\nu_{ij}/T_{0})  }
\end{equation}
where $T_{0}=296\:$K for the HITRAN and HITEMP line lists. Consequently, all intensities of the RNT2017 strong line lists were converted to 296$\:$K, then merged into a global list of $\sim$27 million unique transitions.

The challenge of the second step is to convert the super-line lists into ``effective'' lines that can be used in line-by-line radiative transfer calculations. These are much more flexible than temperature-specific line lists, cross sections or $k$-correlation tables and make the final HITEMP line list more practical. However, the RNT2017 strong line lists are provided at separate temperatures, meaning it is possible for a strong line at $T_{1}$ to be compressed into a super-line at $T_{2}$. Hence, it is also necessary to remove the contribution of the global lines from each super-line list to avoid double counting of individual transitions.

The global line list is calculated at all temperatures given in Tab.~(\ref{tab_ch4_strong_lines}), and the same temperature-dependent thresholds from RNT2017 ($I_{\textrm{\scriptsize{strong}}}(\nu,T)$ and $I_{\textrm{\scriptsize{cut}}}(\nu,T)$) are applied. Considering a transition at $\nu_{1}$ with intensity $I_{1}$ at $T_{1}$, if $I_{\textrm{\scriptsize{strong}}}(\nu_{1},T_{1}) > I_{1} > I_{\textrm{\scriptsize{cut}}}(\nu_{1},T_{1})$ then $I_{1}$ is part of the super-line list at $T_{1}$. The line intensity $I_{1}$ will be included as part of the super-line intensity of the nearest 0.005$\:$cm$^{-1}$ grid point to $\nu_{1}$. The super-line lists are then reprocessed to remove the global strong line contributions. In a small number of cases, the strong line intensity at $T_{1}$ was greater than the corresponding super-line intensity at $T_{1}$. This issue arises because empirical corrections to the RNT2017 strong line lists could not be disentangled from the empirical corrections applied to constituent transitions of each super-line, before they were compressed and the line information lost. It was deemed necessary to remove the line intensity from the super-line lists, even when this intensity had to be removed from a neighboring super-line (to avoid double counting of the strong line intensity). This error is a consequence of attempts to reconstruct the original RNT2017 line list (with 150 billion transitions) prior to compression and can be completely avoided by working with the original line list prior to compression. We strongly recommend that for future investigations, all line lists be retained, prior to the compression into super-lines.

The reprocessed super-line lists are used to produce effective lines that account for the continuum-like absorption of CH$_{4}$. These effective lines must have an effective lower-state energy (allowing conversion of intensities between temperatures) and can then be included with the global line list above. From the intensity ratio of a line as given in Eqn.~(\ref{eqn_int_ratio}), it is possible to determine the $E^{\prime\prime}$ of a transition by comparing the line intensity at different temperatures. Eqn.~(\ref{eqn_int_ratio}) can be rearranged as
\begin{equation}\label{eqn_int_line}
  \textrm{ln}\left[ \frac{S_{ij}(T) Q(T) R(T_{0}) }{ S_{ij}(T_{0}) Q(T_{0}) R(T) } \right] =   \frac{c_{2}E^{\prime\prime}}{T_{0}} - \frac{c_{2}E^{\prime\prime}}{T}
\end{equation}
where $R(T) =  1 - \textrm{exp}(-c_{2}\nu_{ij}/T) $. Thus, a plot of $\textrm{ln}[S_{ij}(T) Q(T) R(T_{0}) / S_{ij}(T_{0}) Q(T_{0}) R(T) ]$ against $-c_{2}/T$ yields the lower-state energy $E^{\prime\prime}$ as the slope. This method has previously been used by \citet{2012ApJ...757...46H,2015ApJ...813...12H} to produce empirical line lists of CH$_{4}$ for high-temperature applications, with a similar two-temperature technique employed by \citet{2012Icar..219..110C} for CH$_{4}$ and included as part of HITRAN2016. This approach is intended to be used for isolated, non-blended transitions with the $E^{\prime\prime}$ provided by a single gradient. However, when applied to blended features, the gradient is determined by the blended feature that dominates the line shape at each temperature \citep{2010ApJ...714..476F}.

\begin{figure*}[p]
\centering
\includegraphics[scale=0.31, trim={0 0 0 0}, clip] {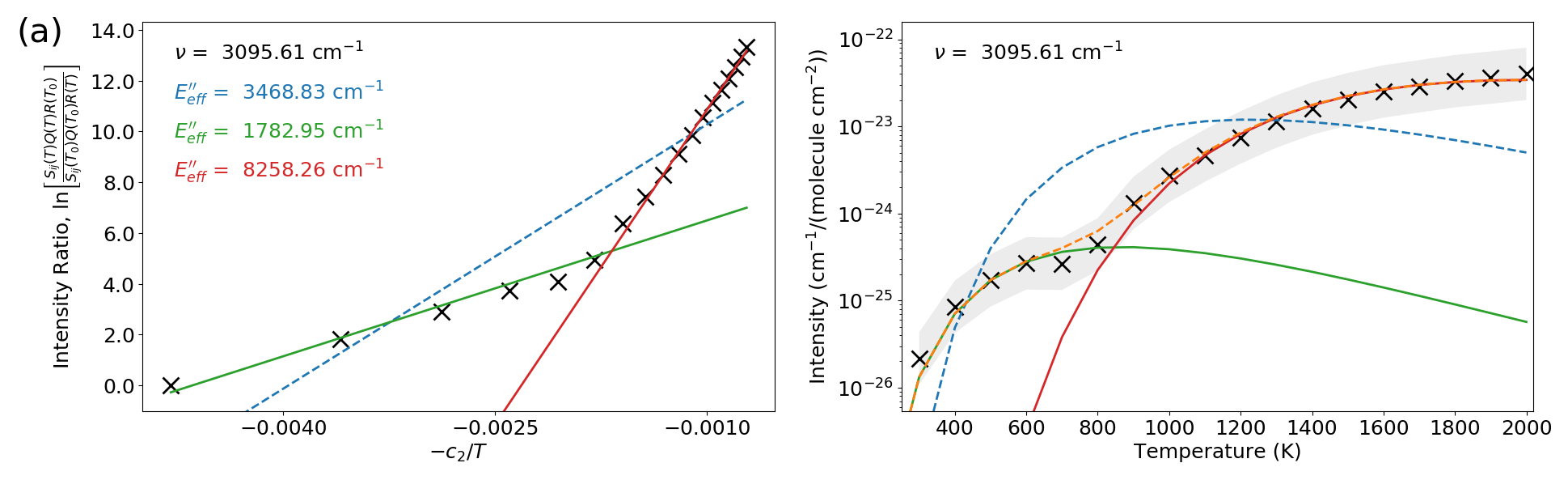} \\
\includegraphics[scale=0.31, trim={0 0 0 0}, clip] {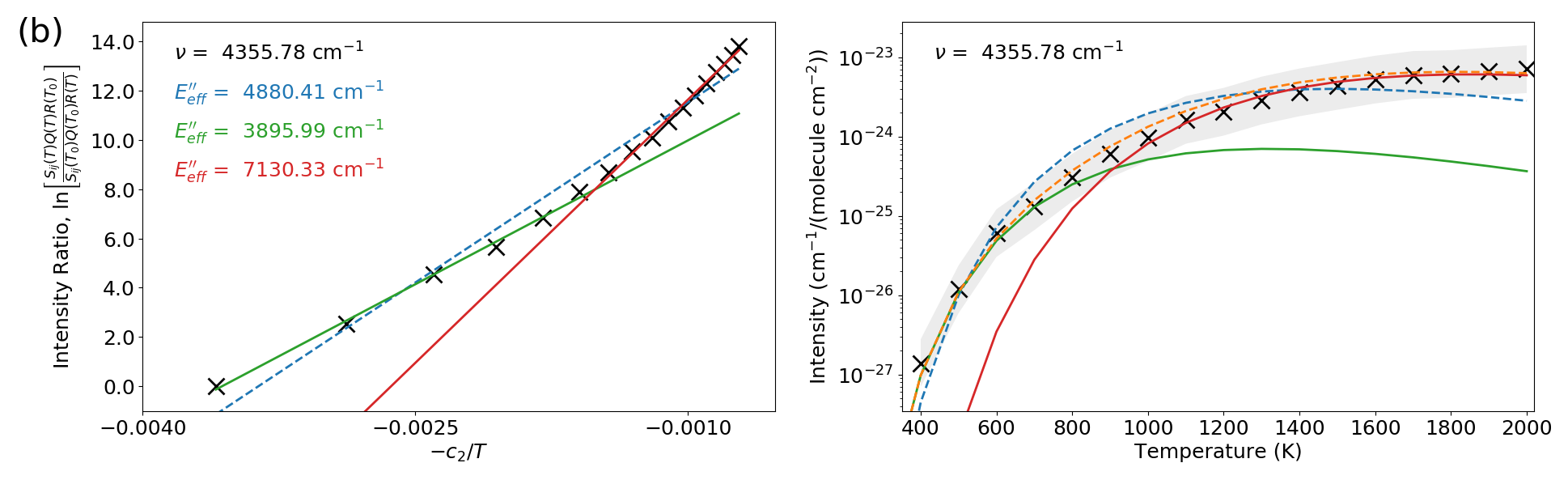} \\
\includegraphics[scale=0.31, trim={0 0 0 0}, clip] {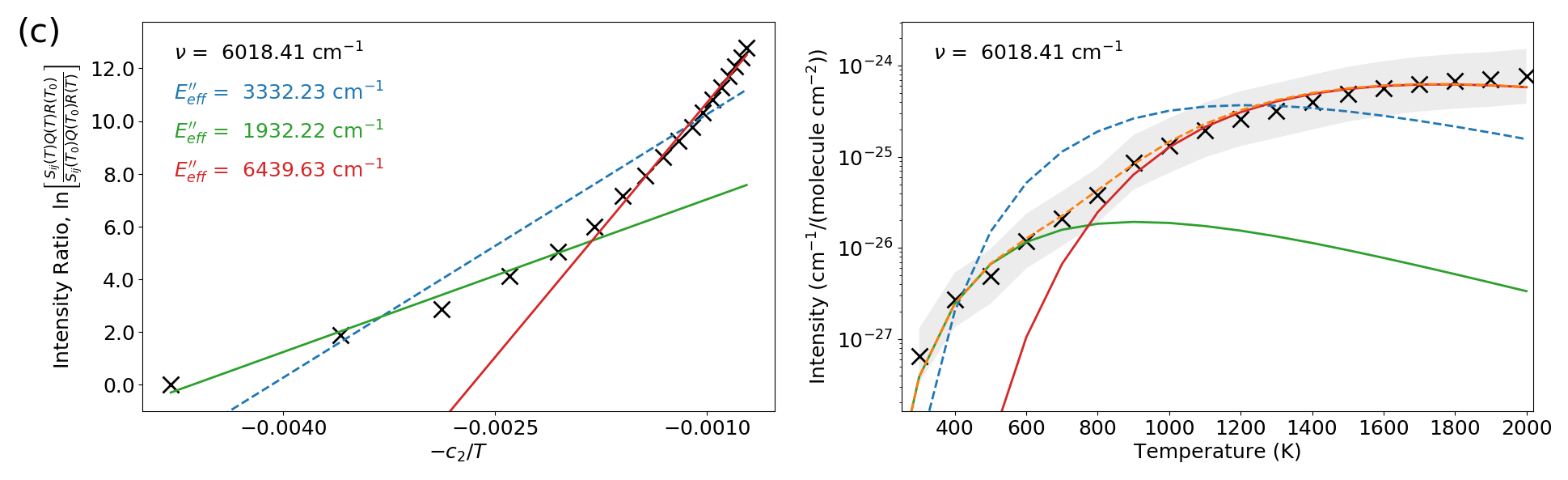} \\
\includegraphics[scale=0.31, trim={0 0 0 0}, clip] {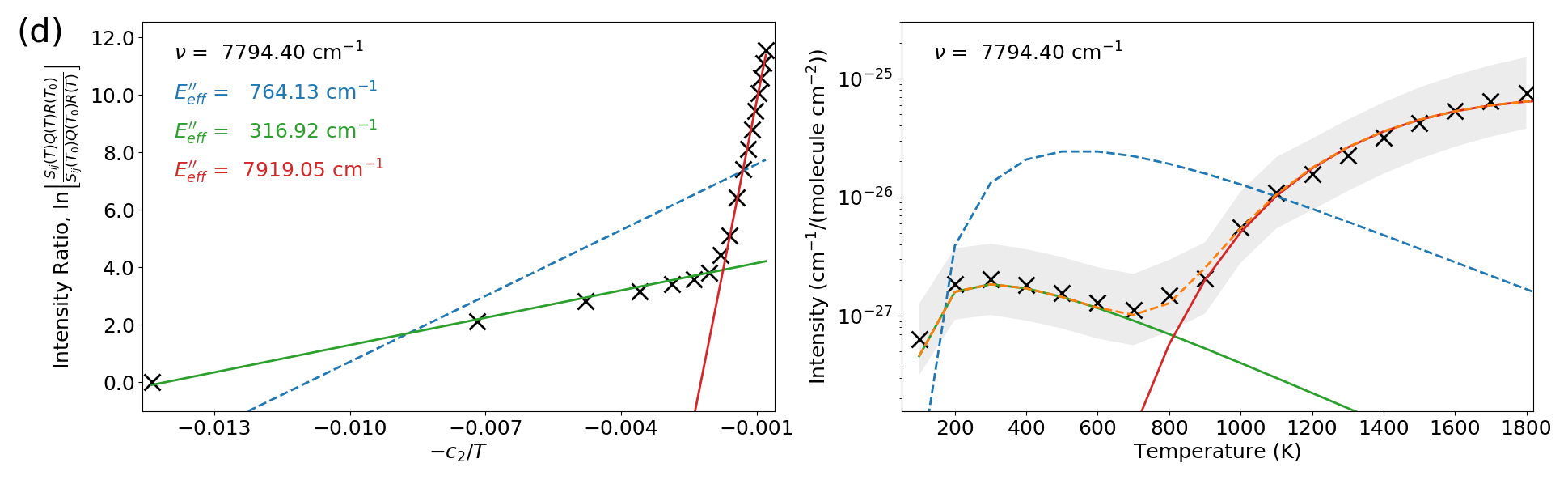}
\caption{Effective lower-state energies ($E_{\textrm{\scriptsize{eff}}}^{\prime\prime}$) have been calculated from the reprocessed super-lines of \citet{2017ApJ...847..105R}. A sample grid point is shown for the pentad (a), octad (b), tetradecad regions (c), and between the icosad and triacontad regions (d). On the left panels, the reprocessed super-line intensity ratios are plotted for the sample grid points ($\nu$ in cm$^{-1}$), using Eqn.~(\ref{eqn_int_line}). The retrieved values of $E_{\textrm{\scriptsize{eff}}}^{\prime\prime}$ are provided (in cm$^{-1}$) for a single line fit (dashed blue) and dual line fit, where the cold and hot component fits are solid green and red lines, respectively. The right panels display the reprocessed super-line intensities as a function of temperature for the same grid points, with the shaded region highlighting an upper/lower bound of a factor of two. In each case, the retrieved values of $E_{\textrm{\scriptsize{eff}}}^{\prime\prime}$ have been used to calculate the intensity contribution from the single line fit (dashed blue) and dual line fits (green and red) at each temperature, with the combined dual line fit given as a dashed orange line.\label{fig_ret_elow}}
\end{figure*}

Applying this technique to the reprocessed super-line lists, it is possible to infer effective lower-state energies, $E_{\textrm{\scriptsize{eff}}}^{\prime\prime}$, for each super-line (i.e.,  at each 0.005$\:$cm$^{-1}$ grid point), such that the intensity at all temperatures can be recovered. In actuality, retrieving a single effective line from each super-line grid point is too simplistic. For example, at 2000~K, $\sim$41~billion weak transitions have been compressed into 1.3~million super-lines: an average of $\sim$31~thousand per super-line. However, in practice the intensity of the super-line appears to be dominated by a single transition or, more likely, the combined intensity of multiple transitions with similar $E^{\prime\prime}$ over a range of temperatures. Hence, it is possible to retrieve an $E_{\textrm{\scriptsize{eff}}}^{\prime\prime}$ of a ``hot'' and ``cold'' component for each 0.005$\:$cm$^{-1}$ super-line grid point.

\begin{figure*}[p]
\centering
\includegraphics[scale=0.42, trim={0 0 0 0}, clip] {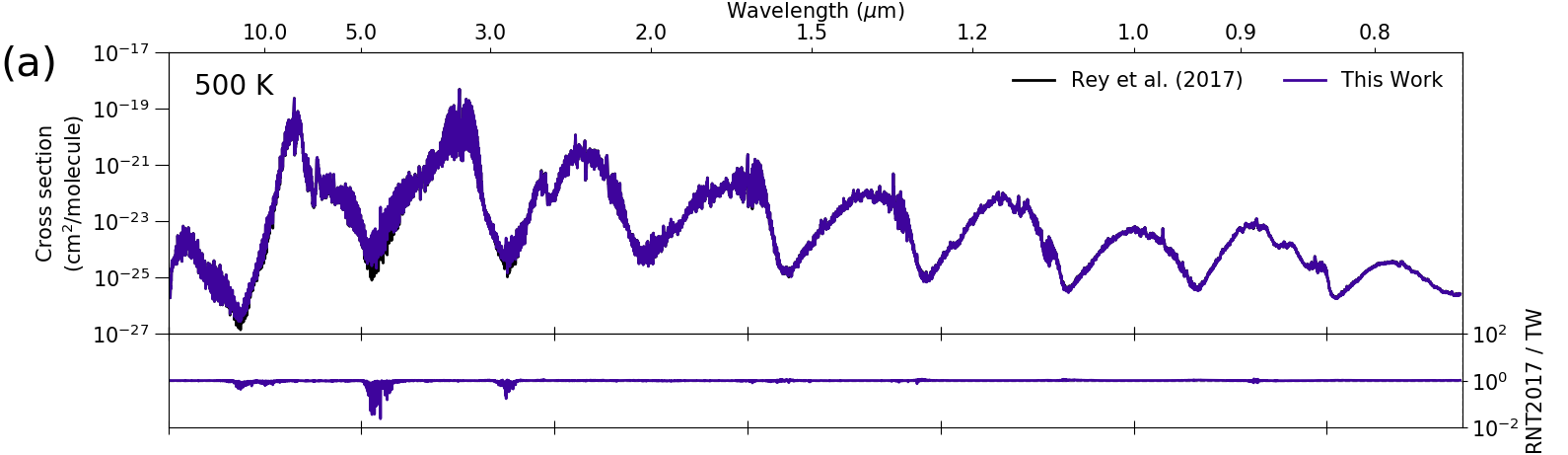} \\
\includegraphics[scale=0.42, trim={0 0 0 0}, clip] {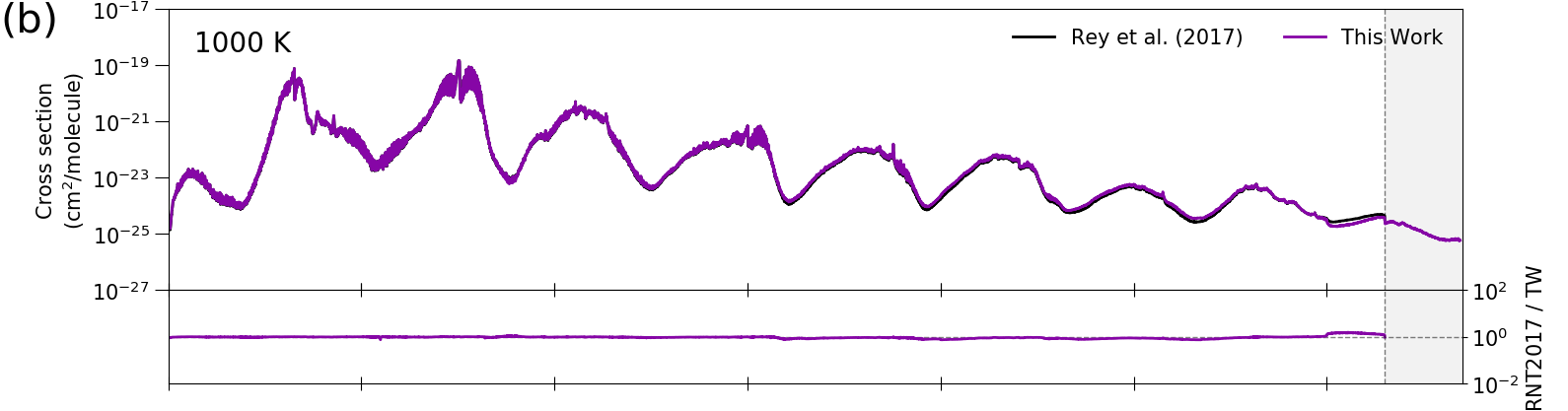} \\
\includegraphics[scale=0.42, trim={0 0 0 0}, clip] {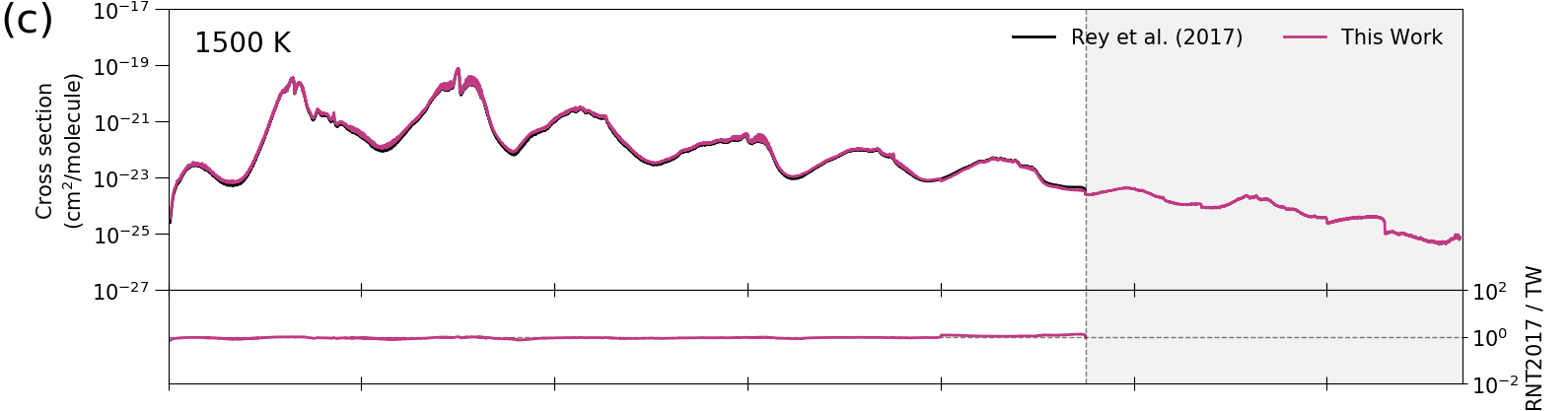} \\
\includegraphics[scale=0.42, trim={0 0 0 0}, clip] {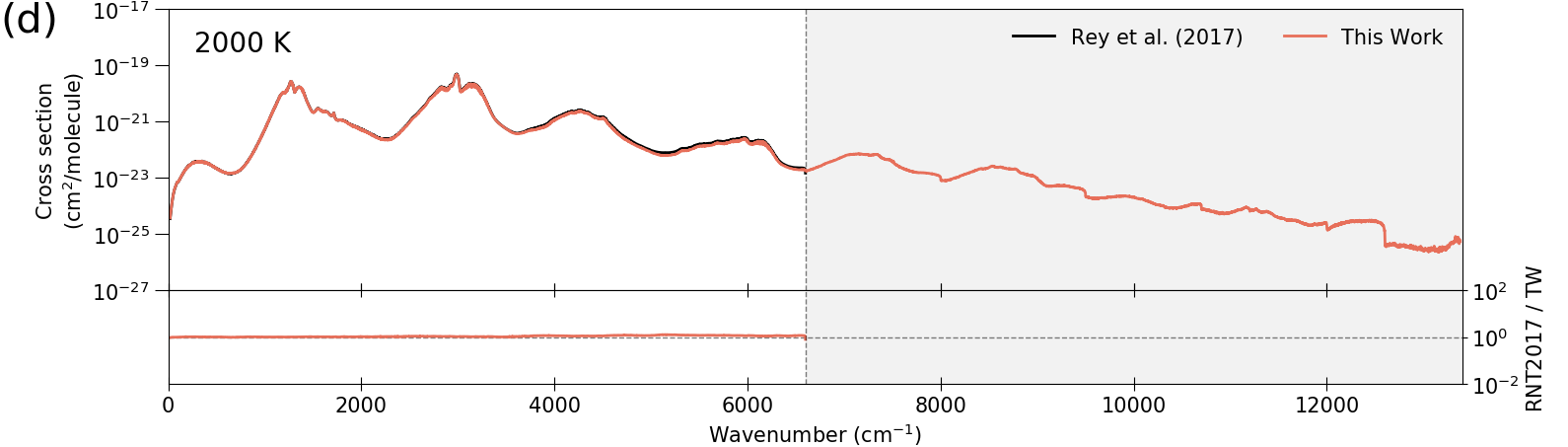}
\caption{Comparisons of the RNT2017 lines lists against the more flexible line list from this work at (a) 500$\:$K, (b) 1000$\:$K, (c) 1500$\:$K, and (d) 2000$\:$K. In each panel, the shaded region indicates the spectral region that is beyond the RNT2017 line lists bounds at each temperature and are therefore not considered complete. These cross sections have been calculated using HAPI \citep{2016JQSRT.177...15K}. \label{fig_v_rnt2017}}
\end{figure*}

The left panels of Fig.~(\ref{fig_ret_elow}) display the intensity ratios against $-c_{2}/T$ from Eqn.~(\ref{eqn_int_line}) for four sample grid points located in the pentad ($P_{2}$), octad ($P_{3}$) and tetradecad ($P_{4}$) regions, and the region between the icosad ($P_{5}$) and triacontad ($P_{6}$). As demonstrated, a single line fit does not reproduce the intensity relationship, with two intersecting gradients clearly observed. On the other hand, a dual line fit is able to account for both gradients extremely well. The right panels of Fig.~(\ref{fig_ret_elow}) display the super-line intensities of the same grid points for increasing temperature. The effective parameters retrieved from the fit in the left panels can be used to calculate the intensity of each effective line for the same temperatures. The temperature range of dominance for the hot and cold components of the dual line fit are most clearly observed in Fig.~(\ref{fig_ret_elow}d), with the combined intensity of both fits matching the grid-point intensities extremely well over several orders of magnitude. The retrieved cold component parameters are sensitive to the minimum temperature at which the super-line grid point is populated (often much higher than 300$\:$K) as well as the crossing point for the two gradients. This resulted in a slight overestimation when calculating the intensity of the effective line at 296$\:$K, $S_{\textrm{\scriptsize{eff}}}$(296$\:$K). An empirical scale factor of 0.8 was applied to $S_{\textrm{\scriptsize{eff}}}$(296$\:$K) for the cold line to mitigate this effect.

A dual line fit was attempted for all super-line grid points, but many grid points were not populated for a sufficient number of temperatures to allow for two separate fits. In these cases a single line fit was used. A small number of grid points contained ``noisy'' intensities, due to reprocessing of the super-line lists, and these fits have been excluded. 

In total, 5,099,138 effective lines have been obtained from the analysis of the reprocessed super-line lists, with an average of 380 effective lines per wavenumber. These have been combined with the global strong line list above to give a single $^{12}$CH$_{4}$ line list of $\sim$32~million lines capable of reproducing the intensities of the strong and super-lines from RNT2017. The effective lines have a special label ``el'' in the assignment part of the resultant line list to emphasize that they do not correspond to an actual transition between $^{12}$CH$_{4}$ energy levels. Since the effective lines do not have rotational quantum assignments, it is not possible to calculate a statistical weight nor Einstein-A coefficient for these lines and consequently these parameters are set to zero.

\subsection{Broadening parameters and HITEMP format} \label{sec:combining:broadening}

Pressure-dependent self-broadening ($\gamma_{\textrm{\scriptsize{self}}}$), air-broadening ($\gamma_{\textrm{\scriptsize{air}}}$) and its temperature dependence ($n_{\textrm{\scriptsize{air}}}$) have been calculated for each strong line based on \citet{2013JQSRT.130..201B}, which describes the CH$_{4}$ line list parameters included in HITRAN2012 \citep{2013JQSRT.130....4R}. The broadening parameters depend on rotational assignments and cannot be directly applied to the effective lines. Instead, values of $\gamma_{\textrm{\scriptsize{self}}} = 0.0680\:$cm$^{-1}$/atm, $\gamma_{\textrm{\scriptsize{air}}} = 0.0519\:$cm$^{-1}$/atm and $n_{\textrm{\scriptsize{air}}} = 0.66$ have been used, based on averaging HITRAN2016 parameters for $^{12}$CH$_{4}$. These effective lines will therefore be indistinguishable from the strong lines when used in line-by-line radiative transfer codes, except for the ``el'' (effective line) identifier as part of the line assignment.  A pressure-dependent line shift has been approximated from line positions as $\delta = - 2\nu \times 10^{-6}\:$cm$^{-1}$/atm.
In the context of high-temperature applications, there is a large room for improvement for these line-shape parameters. For instance, the HITRAN default format allows only temperature dependence for $\gamma_{\textrm{\scriptsize{air}}}$, and using this temperature dependence for $\gamma_{\textrm{\scriptsize{self}}}$ is only an approximate solution. Furthermore, recent works that study the line shape effects over a broad range of temperatures \citep{2018JQSRT.217..440G, Stolarczyk2019} propose the use of a double power law as opposed to a power law with a single exponent. With that being said, \citet{Vispoel2019} recently studied  CH$_{4}$ lines broadened by N$_{2}$ but did not observe a large discrepancy between a single power law and double power law up to 700$\:$K. 
Another consideration for line broadening of CH$_{4}$ is by ``planetary'' gases, including CO$_{2}$, H$_{2}$, He and H$_{2}$O. As previously discussed, HITRAN provides line broadening by CO$_2$, H$_2$, He  and H$_2$O \citep{2016JQSRT.168..193W, 2019...Tan..H2..HIT} for several gases. But for CH$_{4}$, broadening by H$_{2}$O is the only additional perturber currently available \citep{2019...Tan..H2..HIT}. To obtain water-broadened parameters, \citet{2019...Tan..H2..HIT} recommend multiplying $\gamma_{\textrm{\scriptsize{air}}}$ by a single scaling factor of 1.36 and multiplying $n_{\textrm{\scriptsize{air}}}$ by a factor of 1.26. These factors can be applied to the HITEMP line list from this work when doing appropriate calculations. Broadening parameters for other gases will be added to the database in the near future as a response to the increasing amount of relevant experimental and theoretical studies. For instance, \citet{Gharib-Nezhad2019} recently measured broadening of CH$_{4}$ lines by H$_{2}$ over an extended range of temperatures. Finally, the HITRAN database has recently introduced advanced line-shape profiles \citep{Wciso2016}, due to the flexibility offered by the relational database structure. These advanced line shapes can decrease residuals in terrestrial atmospheric spectra to the sub-percent level. While HITEMP line lists could also benefit from their inclusion with respect to high-resolution combustion measurements, the main target of this work is astrophysical applications where such accuracy on the line-shape parameters is not required. 

\subsection{Comparison to RNT2017} \label{sec:combining:rnt2017}

Fig.~(\ref{fig_v_rnt2017}) contains a comparison of calculated absorption cross sections using the total line list from this work against line lists of RNT2017 at four temperatures (combining strong and super-line components). These cross sections are calculated on a fine 0.001$\:$cm$^{-1}$ grid for the full 0 to 13,400$\:$cm$^{-1}$ spectral range with 100~Torr of $^{12}$CH$_{4}$, with calculations performed using HAPI \citep{2016JQSRT.177...15K}. To make the comparisons appropriate, the same broadening and temperature dependence has been applied to the RNT2017 line lists as was used for this work. In all cases, this work is able to reproduce the absorption features seen when using the RNT2017 line lists. For the lowest temperatures, the first three window regions display differences when used at high resolution: a consequence of a slight overestimation of the effective line strengths of the cold lines at 296$\:$K. However, these differences contribute an extremely small amount to the total absorption. Tab.~(\ref{tab_ch4_tw}) includes intensity sums from the single line list of this work ($\Sigma S_{\textrm{\scriptsize{TW}}}(T)$) calculated at the same temperatures as RNT2017. These intensity sums have been compared to those of RNT2017, given in Tab.~(\ref{tab_ch4_strong_lines}). It can be seen that the total intensity for the corresponding wavenumber limits agree with RNT2017 to within 2\% up to 1100$\:$K. This difference increases to a maximum of 6\% at 1500$\:$K, before reducing towards 2000$\:$K. It should be noted that the strong-line sums (or effective/super-line sums) are not directly comparable between this work and RNT2017 because of differences between the number of lines, and therefore intensity, included in each sum (see Sect.~(\ref{sec:combining})). This metric is not necessarily representative of the accuracy of the current work because it is heavily weighted to the intensity sum of the dyad and pentad regions, but these intensity sums can be considered an indicator of the uncertainty of the effective line intensities at each temperature. 

In total, this work requires fewer lines than the RNT2017 line lists, yet is significantly more flexible and able to reproduce the RNT2017 absorption. Cross-sections calculated for this work have been done so using the second generation of HAPI \citep{2016JQSRT.177...15K}, which is available online\footnote{\href{https://github.com/hitranonline/hapi2 }{https://github.com/hitranonline/hapi2}}. The updates to HAPI mean that absorption cross sections calculated from a line list of $\sim$32~million can be processed in approximately 450 seconds on a 12 core 2.6$\:$GHz CPU\footnote{These tests have been performed with line wings set to 25$\:$cm$^{-1}$ for each line; the time may vary depending on the width of line wings taken into account.}.

\begin{deluxetable*}{c|ccc|c}
\tablecaption{Intensity sums for the total $^{12}$CH$_{4}$ line list from this work ($\Sigma S_{\textrm{\scriptsize{TWtot}}}(T)$), along with the strong and effective components. For comparison with Tab.~(\ref{tab_ch4_strong_lines}), these intensity sums have been calculated at the same temperatures. \label{tab_ch4_tw}}
\tablehead{
\colhead{}                      & \colhead{$\Sigma S_{\textrm{\scriptsize{TWstr}}}(T)$\tablenotemark{{\scriptsize a}}, }              &
\colhead{$\Sigma S_{\textrm{\scriptsize{TWeff}}}(T)$\tablenotemark{{\scriptsize a}}, }   & \colhead{$\Sigma S_{\textrm{\scriptsize{TWtot}}}(T)$\tablenotemark{{\scriptsize a}}, }               & \colhead{ $\Sigma S_{\textrm{\scriptsize{TWtot}}}(T)$ $/$  }  \\
\colhead{$T$}                         & \colhead{$\times 10^{-17}$}   &
\colhead{$\times 10^{-17}$}  & \colhead{$\times 10^{-17}$}   & \colhead{  $\Sigma S_{\textrm{\scriptsize{RNTtot}}}(T)$    } \\
\colhead{(K)}                      & \colhead{(cm$^{-1}$/(molecule$\:$cm$^{-2}$))}            &
\colhead{(cm$^{-1}$/(molecule$\:$cm$^{-2}$))}            & \colhead{(cm$^{-1}$/(molecule$\:$cm$^{-2}$))}            & \colhead{(\%)}
}
\startdata
 300 &    1.773  &  0.0000  &  1.773 &  100.0  \\
 400 &    1.776  &  0.0002  &  1.776 &  100.0  \\
 500 &    1.781  &  0.0007  &  1.782 &  100.0  \\
 600 &    1.787  &  0.0023  &  1.789 &  100.0  \\
 700 &    1.794  &  0.0065  &  1.800 &  100.0  \\
 800 &    1.797  &  0.0164  &  1.813 &  100.0  \\
 900 &    1.792  &  0.0370  &  1.829 &  100.3  \\
1000 &    1.777  &  0.0741  &  1.851 &  101.0  \\
1100 &    1.746  &  0.1320  &  1.878 &  101.8  \\
1200 &    1.697  &  0.2113  &  1.908 &  103.2  \\
1300 &    1.631  &  0.3080  &  1.939 &  104.4  \\
1400 &    1.548  &  0.4143  &  1.962 &  105.5  \\
1500 &    1.450  &  0.5204  &  1.970 &  106.0  \\
1600 &    1.343  &  0.6164  &  1.959 &  105.7  \\
1700 &    1.230  &  0.6991  &  1.929 &  105.0  \\
1800 &    1.117  &  0.7622  &  1.879 &  103.5  \\
1900 &    1.003  &  0.7990  &  1.802 &  101.0  \\
2000 &    0.895  &  0.8198  &  1.715 &   97.0  \\
\enddata
\tablenotetext{{\scriptsize a}}{Calculated up to the file limits given in Tab.~(\ref{tab_ch4_strong_lines}) for natural abundance intensities.}
\end{deluxetable*}

For the strong lines that have been empirically corrected by RNT2017, an uncertainty of between $\pm$0.01-0.001$\:$cm$^{-1}$ is obtained, which increases for lines with no empirical corrections.  This uncertainty corresponds to the value reported by \citet{2017ApJ...847..105R} for the complete line lists. On a line-by-line basis, some of the strongest features will have an uncertainty much less than this value, whereas above 6300~cm$^{-1}$ the majority of hot lines have not been catalogued and the positions are not well known. For the strong line intensities, a cautious $\geqq$20\% uncertainty is suggested, although for strong lines within polyads up to $P_{5}$ the intensities can be much more accurate. The uncertainly of the effective line intensities and positions has not been assigned  since the contributing lines are not observed and difficult to quantify. However the comparisons above indicate that these effective lines fall within the uncertainties of the strong lines.

For the remaining line parameters, an uncertainly $\geqq$20\% is given for $\gamma_{\textrm{\scriptsize{self}}}$, $\gamma_{\textrm{\scriptsize{air}}}$ and $n$, and between $\pm$0.01-0.001$\:$cm$^{-1}$ for $\delta$.

To remain consistent with HITRAN, the strong and effective line intensities have been scaled by 0.98827 to account for the natural terrestrial abundance of $^{12}$CH$_{4}$.

\subsection{HITRAN2016 replacements and isotopologue inclusions} 
\label{sec:combining:isos}

Where possible, owing to their high reliability, HITRAN2016 parameters for $^{12}$CH$_{4}$ have been used in place of the RNT2017 values in this work (TW). Matches were identified based on the criteria  $\nu_{\textrm{\scriptsize{TW}}}  = \nu_{\textrm{\scriptsize{HIT2016}}} \pm 0.01\:$cm$^{-1}$, $E_{\textrm{\scriptsize{TW}}}^{\prime\prime} = E_{\textrm{\scriptsize{HIT2016}}}^{\prime\prime} \pm 0.01\:$cm$^{-1}$, $S_{\textrm{\scriptsize{TW}}} = S_{\textrm{\scriptsize{HIT2016}}} \pm 20$~\% and a consistent $J^{\prime\prime}$ between line lists. The wavenumber criteria were relaxed to $\pm$1.0$\:$cm$^{-1}$ for transitions greater than 10,000$\:$cm$^{-1}$ due to a reduced accuracy of the RNT2017 data. A requirement for matching $J^{\prime\prime}$ and $E^{\prime\prime}$ means that regions above 6230$\:$cm$^{-1}$, where assignments are limited or lacking, have very few HITRAN2016 replacements. In total, the $^{12}$CH$_{4}$ line list from this work contains 81,245 lines replaced by HITRAN2016, which amounts to approximately 50\% of the HITRAN2016 $^{12}$CH$_{4}$ line list.

The corresponding theoretical line lists of methane isotopologues is currently not sufficient for high-temperature applications, although  progress is being made towards assignments \citep{2014JChPh.141d4316R, 2015JPCA..119.4763R, 2018JMoSp.351...14K, 2018Icar..303..114R, 2019JQSRT.235..278S}. While it is clear that these line lists are not as complete as $^{12}$CH$_{4}$, their absorption contributes just over 1\% for terrestrial abundances. Therefore at this moment the HITRAN2016 line lists for $^{13}$CH$_{4}$, $^{12}$CH$_{3}$D and $^{13}$CH$_{3}$D have been included with this work (with abundances 0.011103, 6.15751$\times$10$^{-4}$ and 6.91785$\times$10$^{-6}$ respectively).

\begin{deluxetable}{p{4cm}p{3.5cm}}
\tablecaption{A summary of the CH$_{4}$ line list for HITEMP. \label{tab_hitemp}}
\tablehead{
\colhead{Item}      & \colhead{ Details }              
}
\startdata
Isotopologues included\tablenotemark{{\scriptsize a}}  & $^{12}$CH$_{4}$, $^{13}$CH$_{4}$, $^{12}$CH$_{3}$D, $^{13}$CH$_{3}$D  \\
Total number of lines        &  31,880,412 \\
Proportion of effective lines    &  16.0\%  \\  
$\nu_{max}$($T$)\tablenotemark{{\scriptsize b}}    &    13,400$\:$cm$^{-1}$ ($>$746$\:$nm)  \\
$T_{max}$($\nu$)\tablenotemark{{\scriptsize c}}       &  2000$\:$K \\
$T_{0}$ & Parameters are provided at 296$\:$K  \\
$S(T_{0})$ & Intensities are scaled for natural abundances and given at $T_{0}$ (see text)  \\
\enddata
\tablenotetext{{\scriptsize a}}{See Fig.~(\ref{fig_total_list}) for number of lines per isotopologue.}
\tablenotetext{{\scriptsize b}}{Dependant on the temperature coverage given in Tab.~(\ref{tab_ch4_strong_lines}).}
\tablenotetext{{\scriptsize c}}{Dependant on the spectral range given in Tab.~(\ref{tab_ch4_strong_lines}).}
\end{deluxetable}

\begin{figure}[t!]
\centering
\includegraphics[scale=0.22, trim={0 0 0 0}, clip] {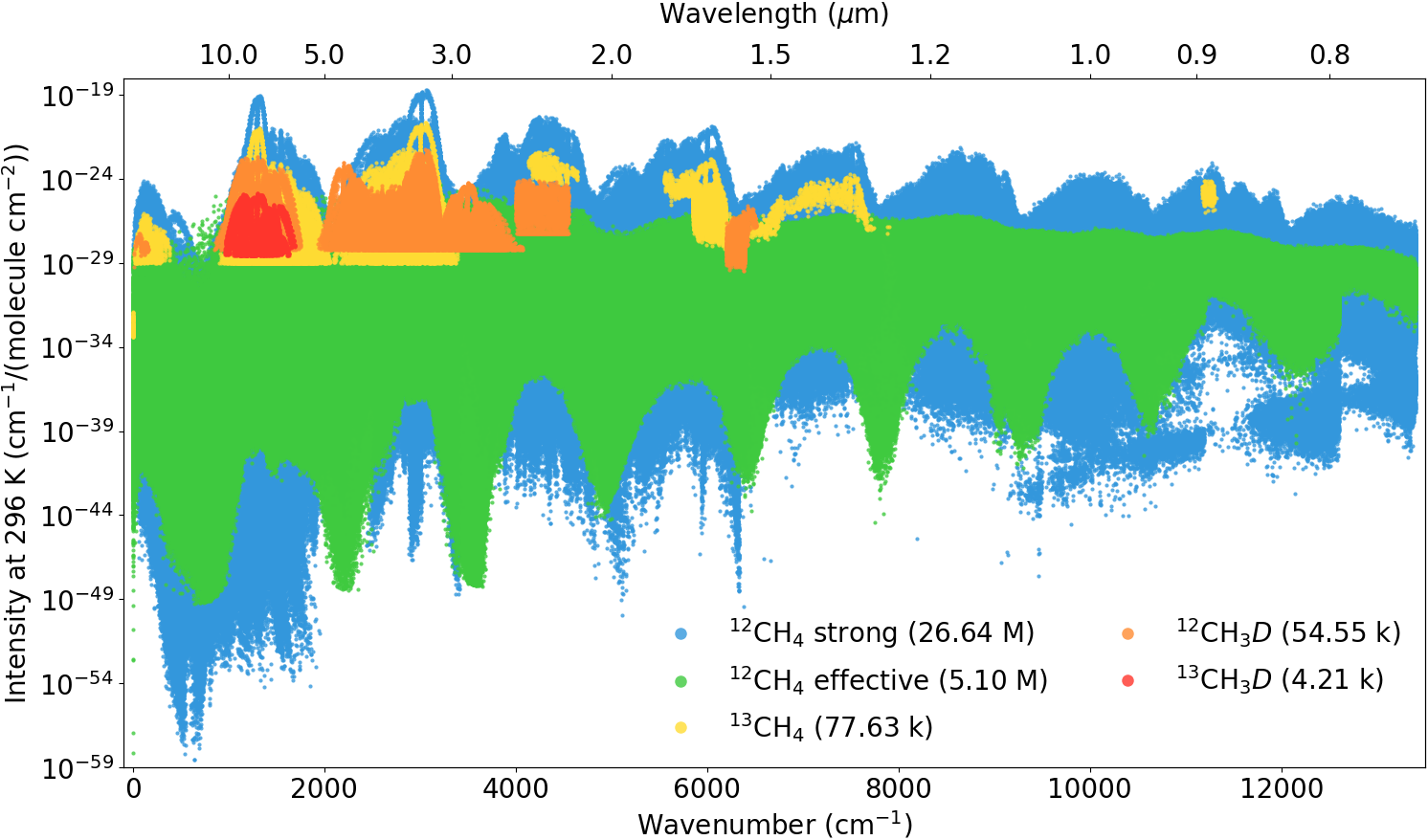}
\caption{The number and coverage of all CH$_{4}$ isotopologues included as part of this work towards a line list for HITEMP, with line intensities plotted at 296$\:$K. The strong and effective lines of $^{12}$CH$_{4}$ have been indicated. \label{fig_total_list}}
\end{figure}

All 31,880,412 CH$_{4}$ lines included in this work for HITEMP are displayed as an overview in Fig.~(\ref{fig_total_list}). The corresponding number of lines for each isotopologue is indicated, with lines of $^{12}$CH$_{4}$ separated into strong and effective components. A summary of the CH$_{4}$ line list for HITEMP is given in Tab.~(\ref{tab_hitemp}).


\section{Room temperature comparisons} \label{sec:tempcomp:low}

The resulting HITEMP line list from this work can be compared to high-resolution absorption cross-sections of HITRAN2016 at 296~K, calculated using the HAPI routines \citep{2016JQSRT.177...15K}. These comparisons can test the accuracy of the known line positions as well as the validity of unassigned features. 

\begin{figure*}[t!]
\centering
\includegraphics[scale=0.50, trim={0 0 0 1cm}, clip] {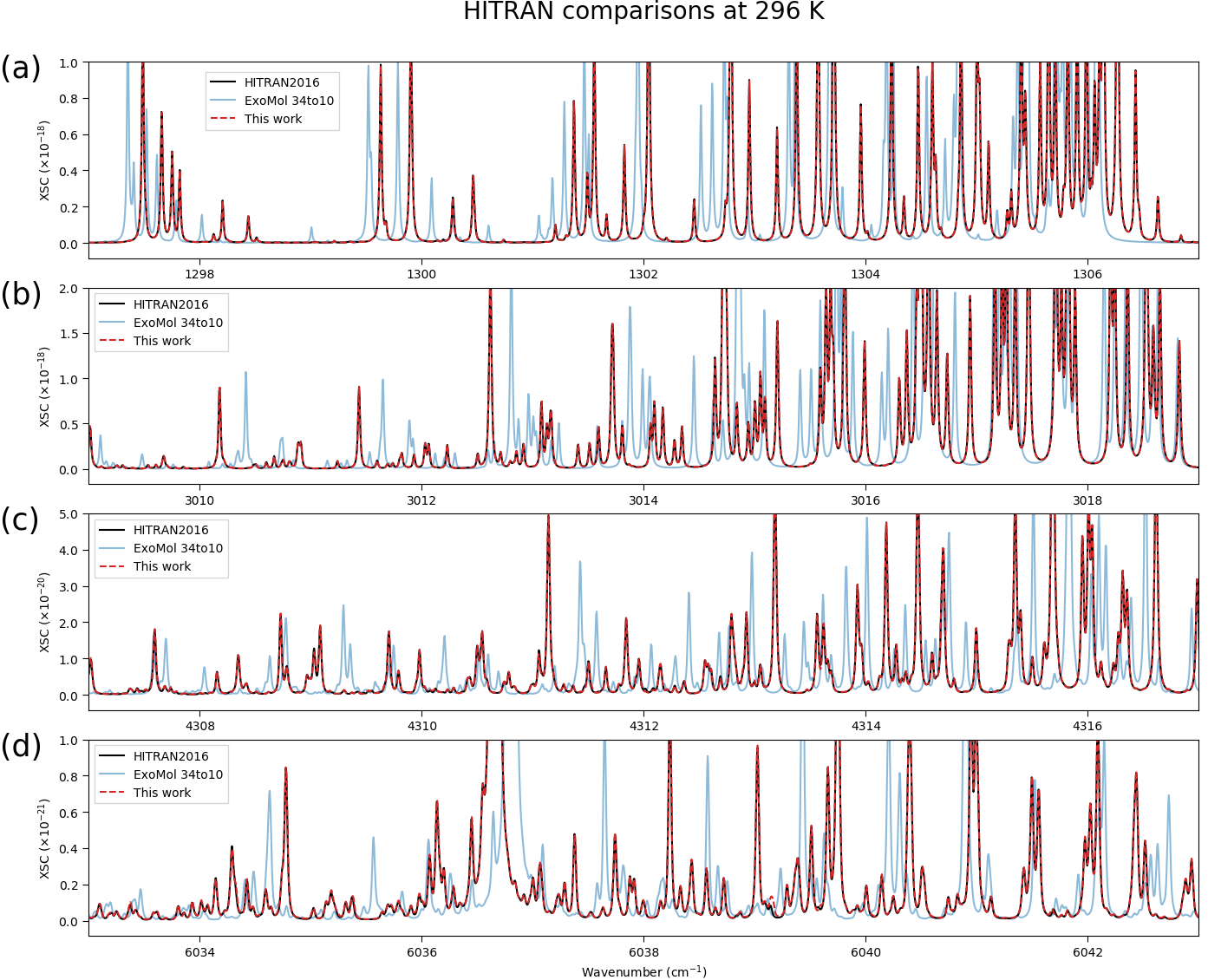}
\caption{Absorption cross sections (in cm$^{2}$/molecule) from HITRAN2016 (black) calculated at a resolution of 0.02$\:$cm$^{-1}$, 296$\:$K and 100$\:$Torr of $^{12}$CH$_{4}$, compared to this work (dashed red) and ExoMol 34to10 (blue). Each panel contains a 10$\:$cm$^{-1}$ region of the dyad (a), pentad (b), octad (c) and tetradecad (d) regions. \label{fig_hitran}}
\end{figure*}

Fig.~(\ref{fig_hitran}) details these comparisons for a 10~cm$^{-1}$ portion of the dyad ($P_{1}$), pentad ($P_{2}$), octad ($P_{3}$) and tetradecad ($P_{4}$) regions at 296$\:$K. For comparison, a cross section calculated using the ExoMol 34to10 line list for the same conditions is also included. In each of these polyads regions, this work is able to replicate all features remarkably well with similar results seen across each band. Only small differences are seen on this scale, with a large number of the strongest lines having positions and intensities identical to HITRAN2016, due to consistent assignments below 6300$\:$cm$^{-1}$.

\begin{figure*}[t!]
\centering
\includegraphics[scale=0.50, trim={0 0 0 1cm}, clip] {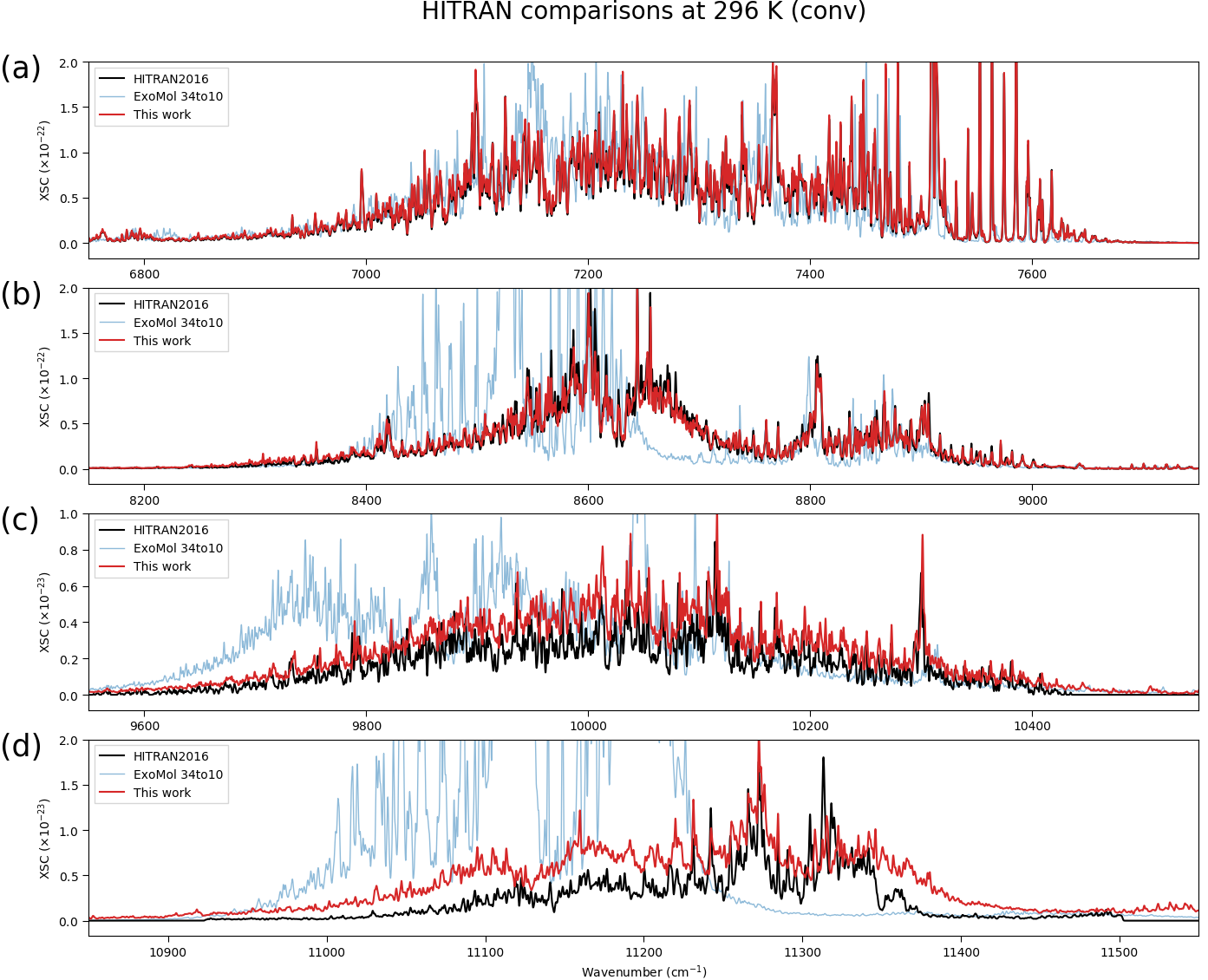}
\caption{Absorption cross sections (in cm$^{2}$/molecule) from HITRAN2016 (black) calculated at 296$\:$K for 100$\:$Torr of $^{12}$CH$_{4}$ and convolved to a resolution of 1.0$\:$cm$^{-1}$, compared to this work (dashed red) and ExoMol 34to10 (blue). Each panel contains an overview of the icosad (a), triacontad (b), tetracontad (c) and pentacontakaipentad (d). \label{fig_hitran_conv}}
\end{figure*}

Fig.~(\ref{fig_hitran_conv}) shows a similar comparison as Fig.~(\ref{fig_hitran}), but for the icosad ($P_{5}$), triacontad ($P_{6}$), tetracontad ($P_{7}$) and pentacontakaipentad ($P_{8}$) regions at 296$\:$K. In this case the absorption cross sections have been convolved to a resolution of 1$\:$cm$^{-1}$ to account for the expected lower accuracy in line positions, and the whole band is displayed. Again, the comparisons between this work and HITRAN2016 for $P_{5}$ and $P_{6}$ are almost identical. Here, the lack of assignments in HITRAN2016 means that the majority of features in this work are provided by strong lines of RNT2017. It is also clear that the ExoMol 34to10 line list has noticeable differences compared to HITRAN2016 and this work at 296~K for these bands. Larger discrepancies between this work and HITRAN2016 are seen for $P_{7}$ and $P_{8}$. However in general the overall structure of each band is maintained, which is not seen for ExoMol 34to10. For the line positions beyond 12,000$\:$cm$^{-1}$, the accuracy of this work is known to be insufficient and differences are observed when compared to low-resolution absorption coefficient band models of \citep{2010Icar..205..674K}.

\begin{figure*}[t!]
\centering
\begin{tabular}{c}
\includegraphics[scale=0.42, trim={0 0 0 0}, clip] {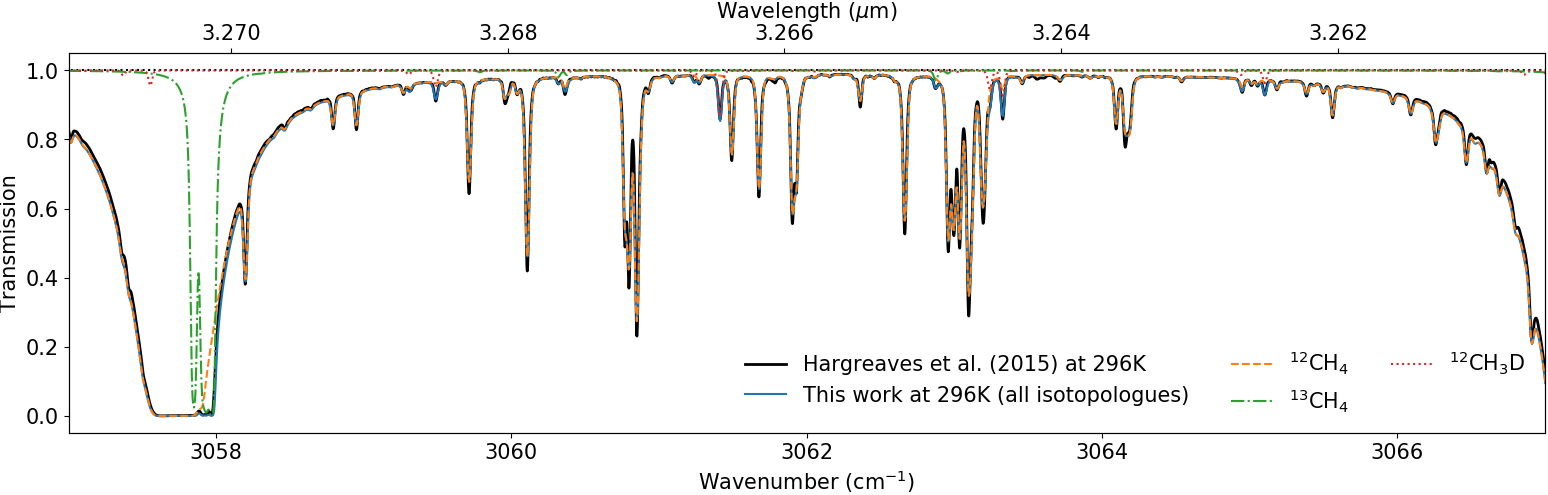}  \\
\includegraphics[scale=0.42, trim={0 0 0 0}, clip] {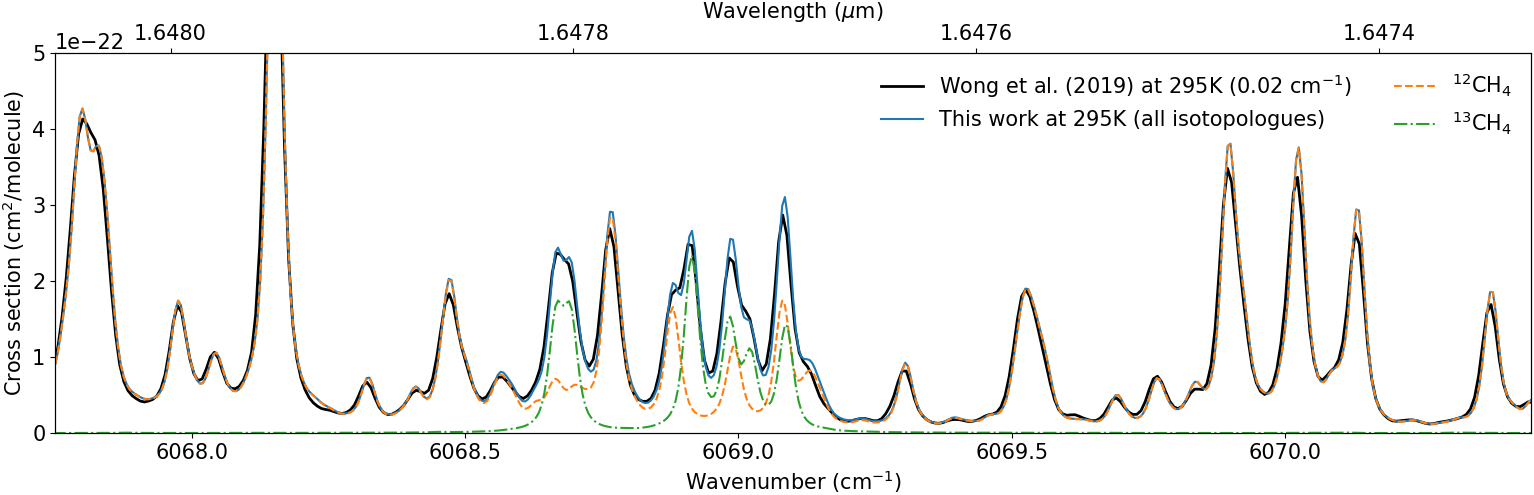}
\end{tabular}
\caption{(a) An experimental transmission spectrum at 296~K from \citet{2015ApJ...813...12H} near 3060$\:$cm$^{-1}$, compared to this work. (b) An experimental absorption cross section at 295$\:$K from \citet{2019ApJS..240....4W} near 6070$\:$cm$^{-1}$, compared to this work. In both panels, the contribution of each isotopologue has been indicated, with only those isotopologues that have lines within the corresponding spectral regions (of this work) shown. \label{fig_isos}}
\end{figure*}

The experimental measurements of \citet{2015ApJ...813...12H} and \citet{2019ApJS..240....4W} contain spectra of CH$_{4}$ with terrestrial abundances. The room-temperature  observations can be used to validate the inclusion of CH$_{4}$ isotopologues from HITRAN2016 into the high-temperature line list for this work. The upper panel of Fig.~(\ref{fig_isos}) includes a CH$_{4}$ transmission spectrum recorded at 296~K covering the pentad region near 3060$\:$cm$^{-1}$. Here, the absorption features of $^{12}$CH$_{4}$, $^{13}$CH$_{4}$ and $^{12}$CH$_{3}$D can be seen with terrestrial abundances. In addition, the lower panel displays an absorption cross section at 295$\:$K for the tetradecad region near 6070$\:$cm$^{-1}$, and again the contribution of the $^{12}$CH$_{4}$ and $^{13}$CH$_{4}$ absorption is clearly observed.

\section{High-temperature comparisons} \label{sec:tempcomp:high}

There are limited spectroscopic observations of CH$_{4}$ at high temperatures. Nevertheless, the RNT2017 line lists available via the last update of the TheoReTS database has previously been validated against high-temperature experimental observations particularly in the frame of the e-PYTHEAS project \citep{2017e-PYTHEAS}) aimed at  astrophysical exoplanetary applications. This included DAS laser absorption spectroscopy experiments at 1000~K in the the region near 6000~cm$^{-1}$ by the Grenoble University group \citep{2018JQSRT.215...59G} and emission spectroscopy experiments at $\sim$1300$\:$K near 3000$\:$cm$^{-1}$ by the Rennes University group \citep{2018JChPh.148m4306A, 2018Amyay.erratum, 2019RScI...90i3103G}. Comparisons with FTS measurement at Norfolk University over a larger spectral range (5600-9000$\:$cm$^{-1}$) were reported as an absorption cross-section atlas at eight temperatures from 300$\:$K to 1000$\:$K \citep{ 2019ApJS..240....4W}.  The TheoReTS calculations provided the best agreement versus these experiments with respect to all other available theoretical lists at elevated temperatures. This was the reason to combine the RNT2017 \textit{ab initio} data with HITRAN2016 in order to construct the new CH$_{4}$ HITEMP database in a user-friendly unified format.  The concept of the present HITEMP work is different from that of TheoReTS due to the modeling of the continuum-like spectral features. Therefore it is necessary to include additional comparisons to validate the line list produced for this work.

High-temperature comparisons have been made to experimental observations of $P_{2}$ and $P_{3}$ from \citet{2015ApJ...813...12H} and $P_{4}$, $P_{5}$ and $P_{6}$ from \citet{2019ApJS..240....4W}. Emission spectra for $P_{1}$ have been measured by \citet{2012ApJ...757...46H}; however self-absorption effects make them inappropriate for comparisons due to unreliable line intensities. The comparisons included here are intended to give a representative overview of the performance of this work at high temperatures.

\begin{figure*}[p]
\centering
\begin{tabular}{c}
\includegraphics[scale=0.36, trim={0 0 0 0}, clip] {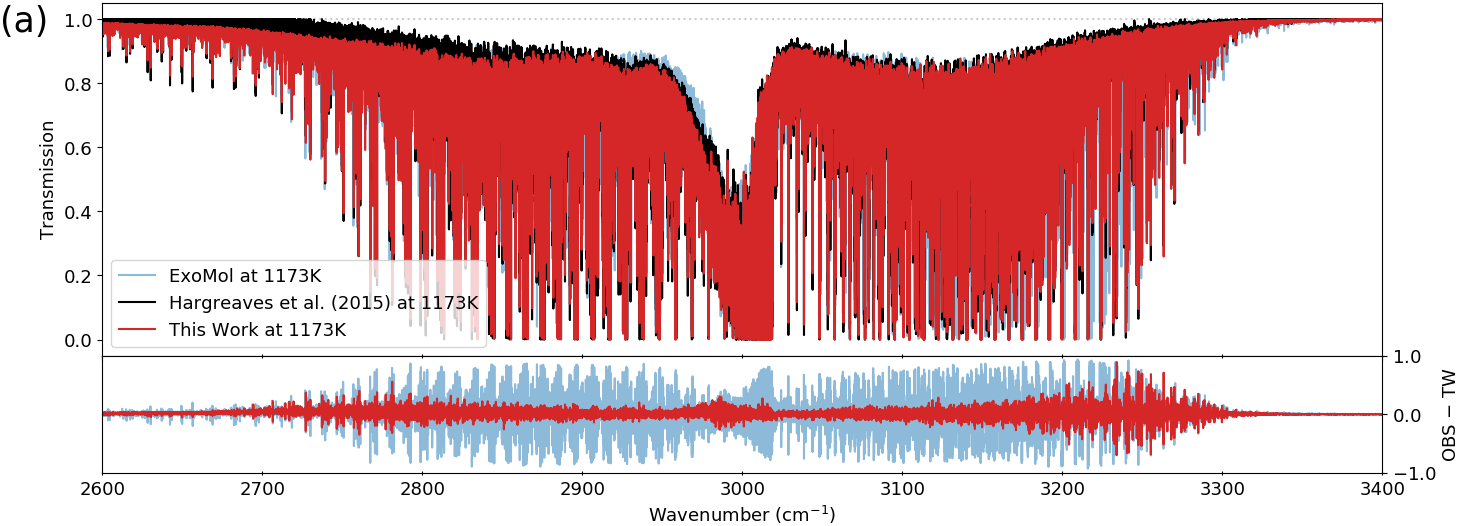} \\
\includegraphics[scale=0.36, trim={0 0 0 0}, clip] {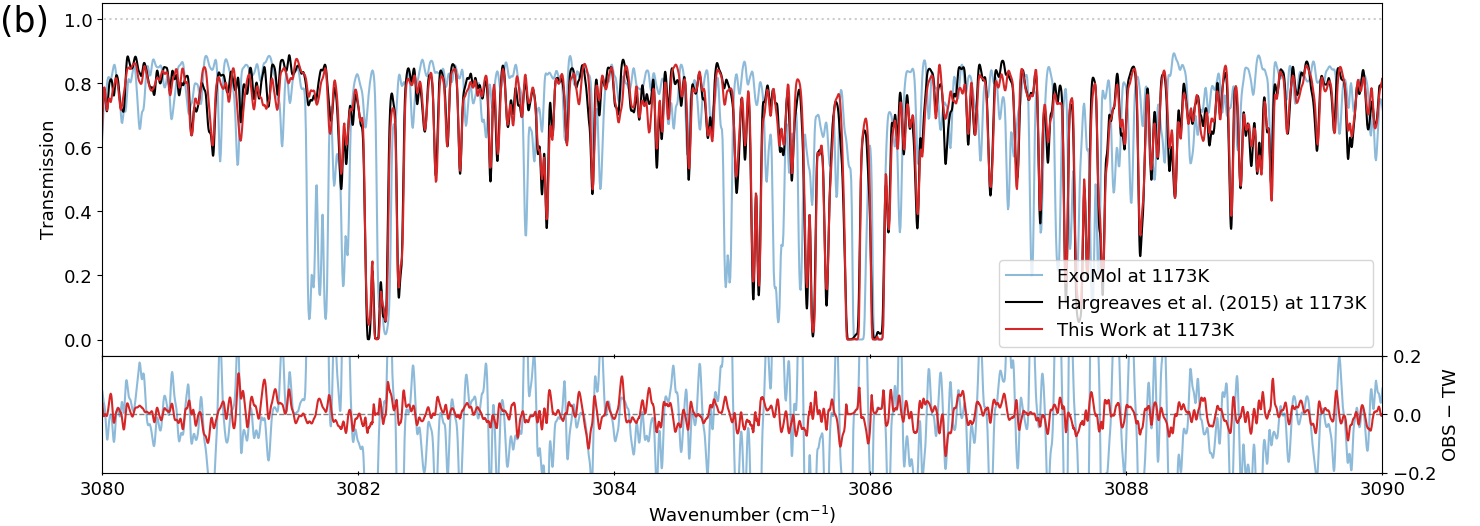} \\
\includegraphics[scale=0.36, trim={0 0 0 0}, clip] {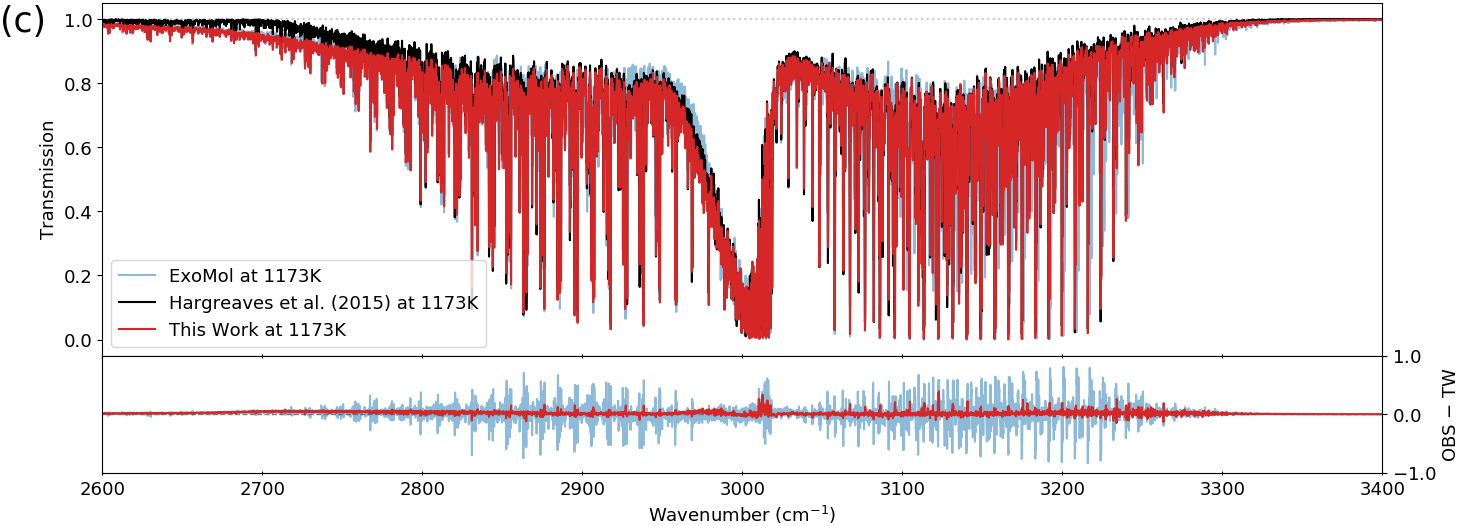} \\
\includegraphics[scale=0.36, trim={0 0 0 0}, clip] {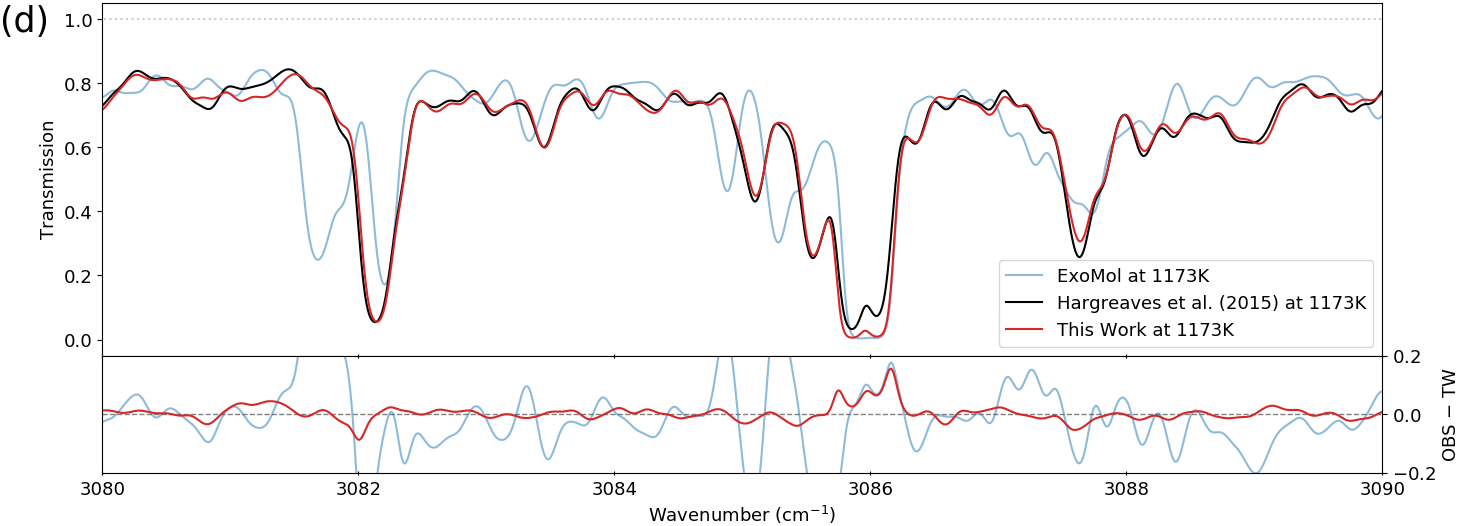}
\end{tabular}
\caption{Comparisons with transmission spectra of \citet{2015ApJ...813...12H} for the pentad region at 1173$\:$K, with 60$\:$Torr of $^{12}$CH$_{4}$ and a path length of 50$\:$cm. Panel (a) displays the full band at a resolution of 0.015$\:$cm$^{-1}$, with (b) showing a zoomed in feature. Panel (c) details the same as (a) but with a convolved resolution of 0.15$\:$cm$^{-1}$, the same zoomed feature is given in (d). In all panels the observations are given in black, this work red, and ExoMol in blue. \label{fig_pentad}}
\end{figure*}

\begin{figure*}[p]
\centering
\begin{tabular}{c}
\includegraphics[scale=0.36, trim={0 0 0 0}, clip] {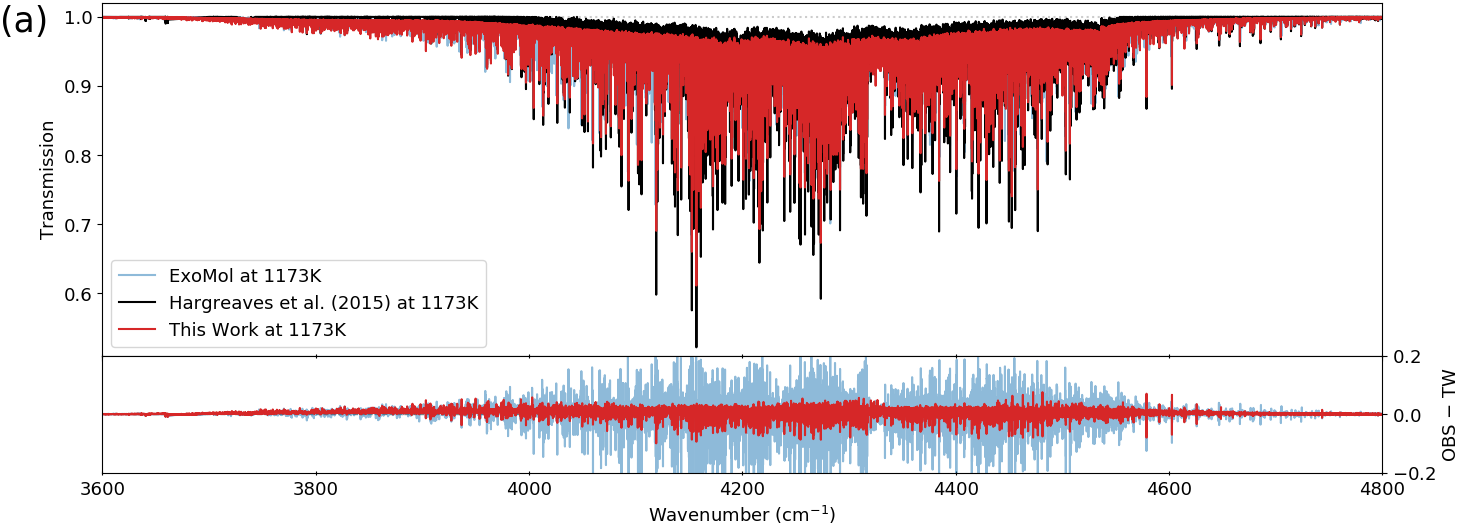} \\
\includegraphics[scale=0.36, trim={0 0 0 0}, clip] {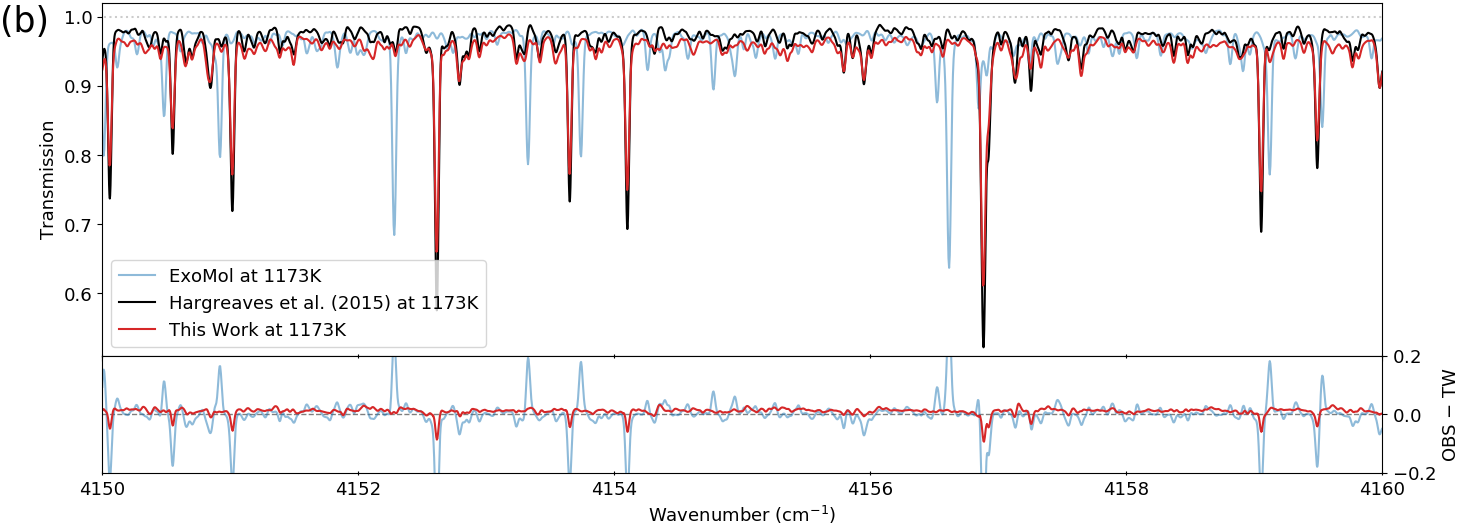} \\
\includegraphics[scale=0.36, trim={0 0 0 0}, clip] {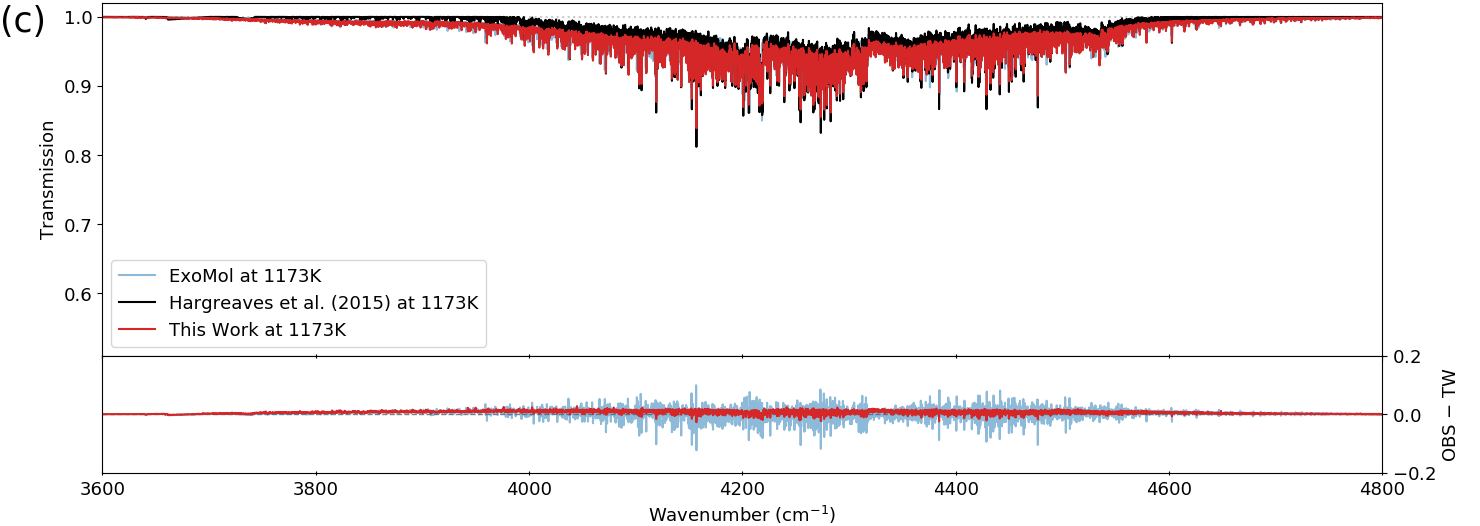} \\
\includegraphics[scale=0.36, trim={0 0 0 0}, clip] {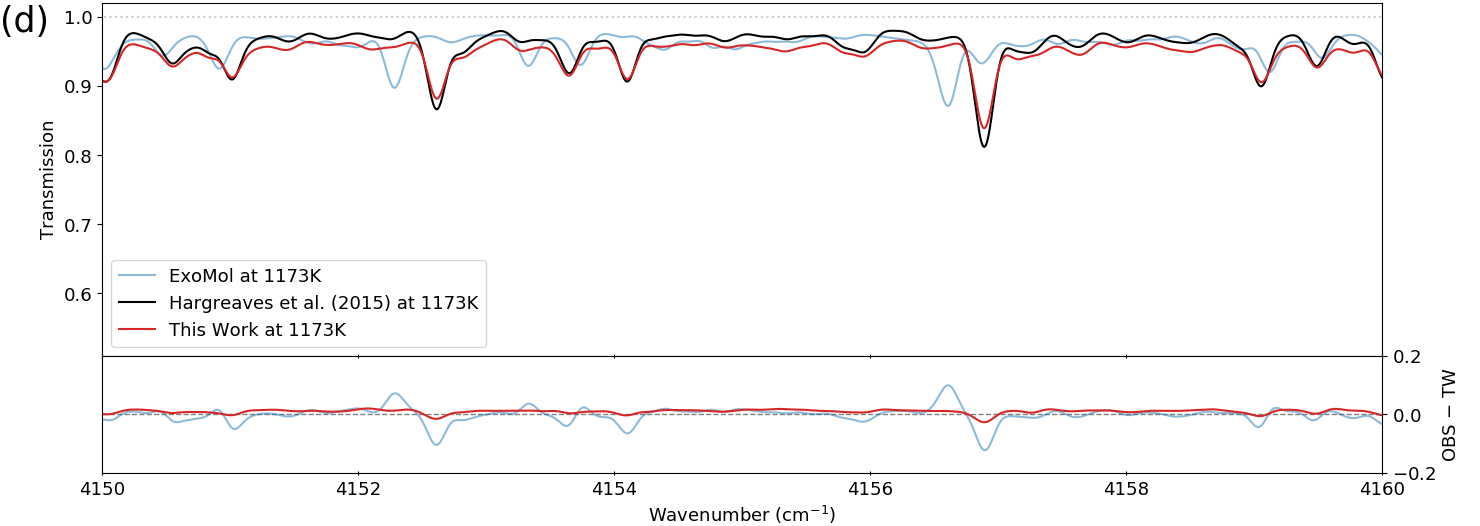}
\end{tabular}
\caption{Comparisons with transmission spectra of \citet{2015ApJ...813...12H} for the octad region at 1173$\:$K, with 60$\:$Torr of $^{12}$CH$_{4}$ and a path length of 50$\:$cm. Panel (a) displays the full band at a resolution of 0.015$\:$cm$^{-1}$, with (b) showing a zoomed in feature. Panel (c) details the same as (a) but with a convolved resolution of 0.15~cm$^{-1}$, the same zoomed feature is given in (d). In all panels the observations are given in black, this work red, and ExoMol in blue.  \label{fig_octad}}
\end{figure*}

Fig.~(\ref{fig_pentad}) and Fig.~(\ref{fig_octad}) compare the observations of \citet{2015ApJ...813...12H} at 1173$\:$K (the maximum temperature observed) with a transmission spectrum calculated using this work and that of ExoMol 34to10. In each calculation, the experimental conditions of 60~Torr of CH$_{4}$ was used with a path length of 50$\:$cm. Panels (a) and (b) compare at the experimental resolution of 0.015$\:$cm$^{-1}$ with good overall agreement to this work for the majority of the band. The line position accuracy decreases for high-$J$ lines towards the edge of the band. Panels (c) and (d) show the same spectra, but have now been convolved to a resolution of 0.15$\:$cm$^{-1}$. In this case the residuals throughout the band have been significantly reduced (except for strong lines that were saturated in the experimental observations and do not convolve correctly), which indicates a larger uncertainly for these high-$J$ lines  as they are not included in HITRAN2016. For comparison, a cross section calculated using the ExoMol 34to10 line list has again been included to demonstrate differences at high temperature. The same comparisons were made for all temperatures available from \citet{2015ApJ...813...12H}, with similar results. At lower temperatures, the line position differences are less significant and this work is able to model the experimental observations to a higher resolution. It is clear that across both polyads, the residuals are much larger for ExoMol 34to10 and line positions are noticeably shifted.  These differences are important to note for high-resolution applications (e.g., cross correlation of exoplanet spectra), which rely on accurate line positions. Hence, this work (and HITEMP) provides the most accurate high-temperature line list available for simulating spectra for these polyad regions up to 1173$\:$K. When using this line list, care should be taken with regards to the accuracy of line positions. As the temperature increases, the intensity of high-$J$ lines and hot bands increase, which consequently reduces the overall accuracy of line positions to $\sim$0.15$\:$cm$^{-1}$ at 1173$\:$K (i.e., $R \sim 20,000$ for the pentad and $R \sim 28,000$ for the icosad). It can be expected that this uncertainty will increase for higher temperatures.

\begin{figure*}[p]
\centering
\begin{tabular}{c}
\includegraphics[scale=0.42, trim={0 0 0 0}, clip] {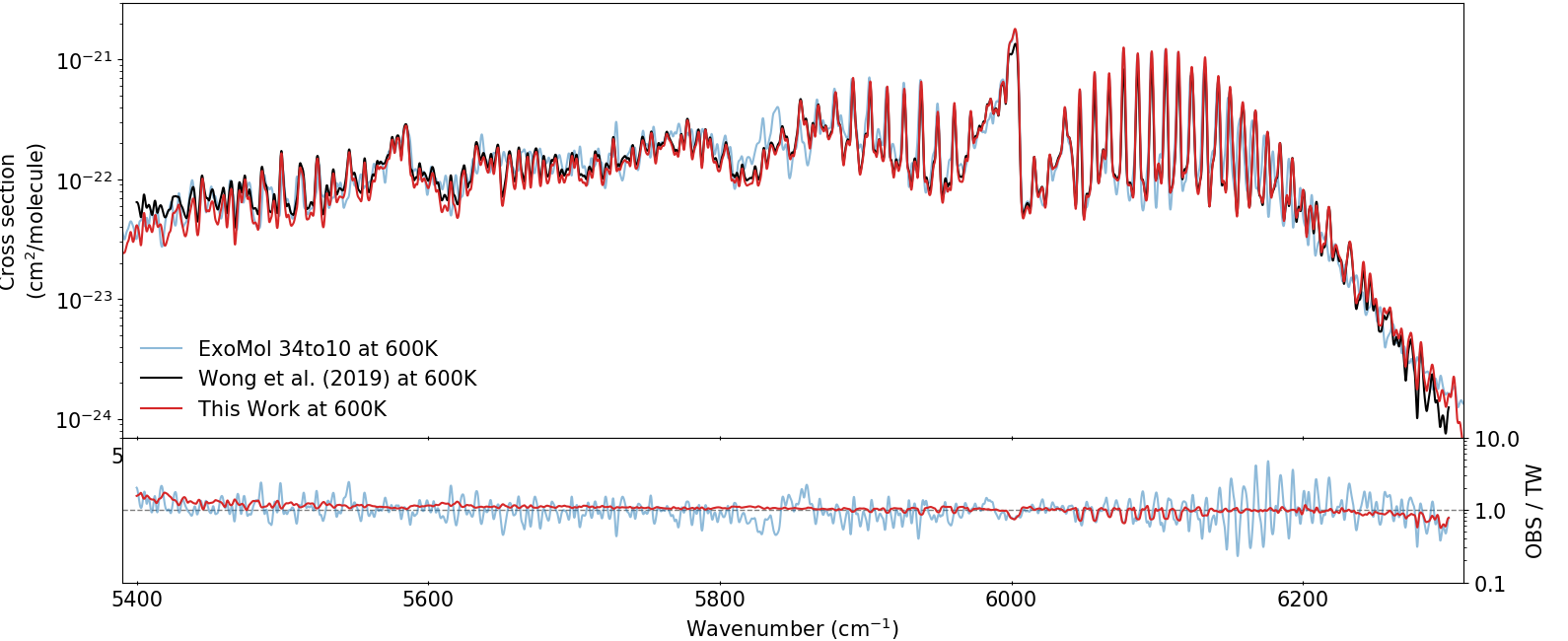} \\
\includegraphics[scale=0.42, trim={0 0 0 0}, clip] {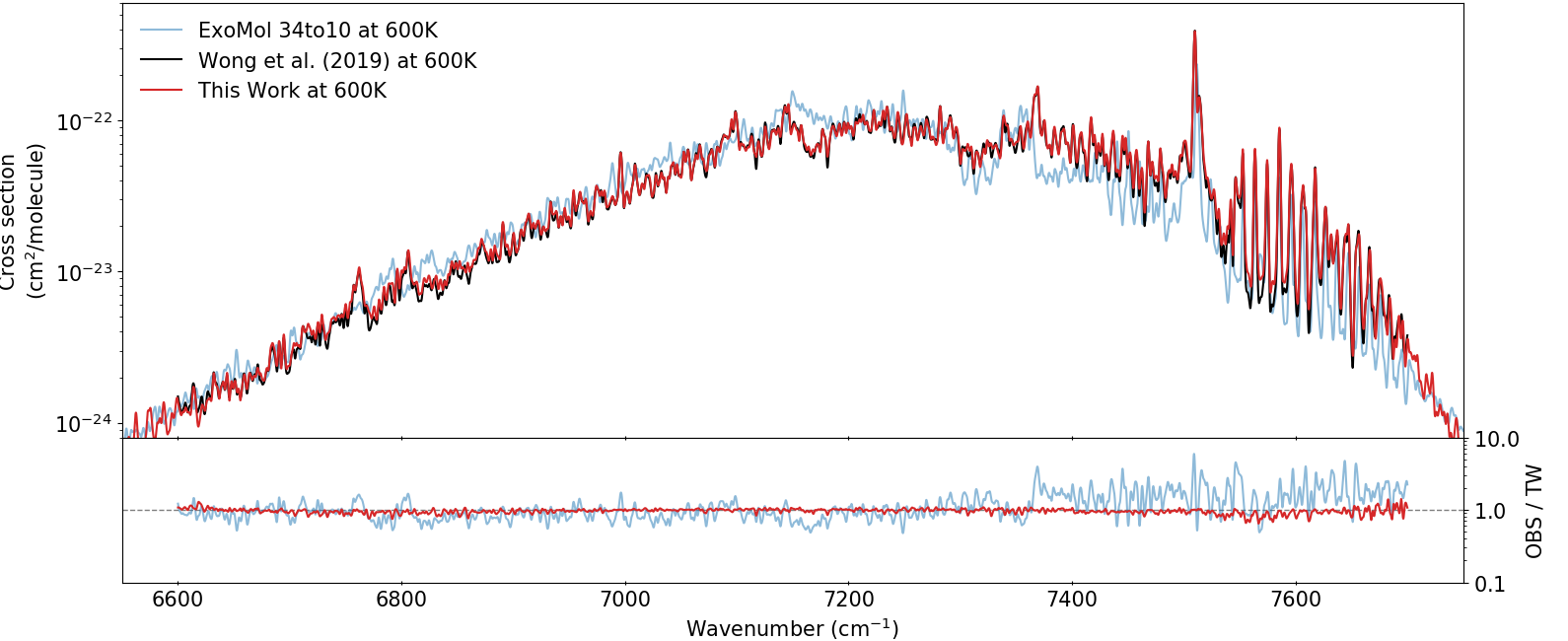} \\
\includegraphics[scale=0.42, trim={0 0 0 0}, clip] {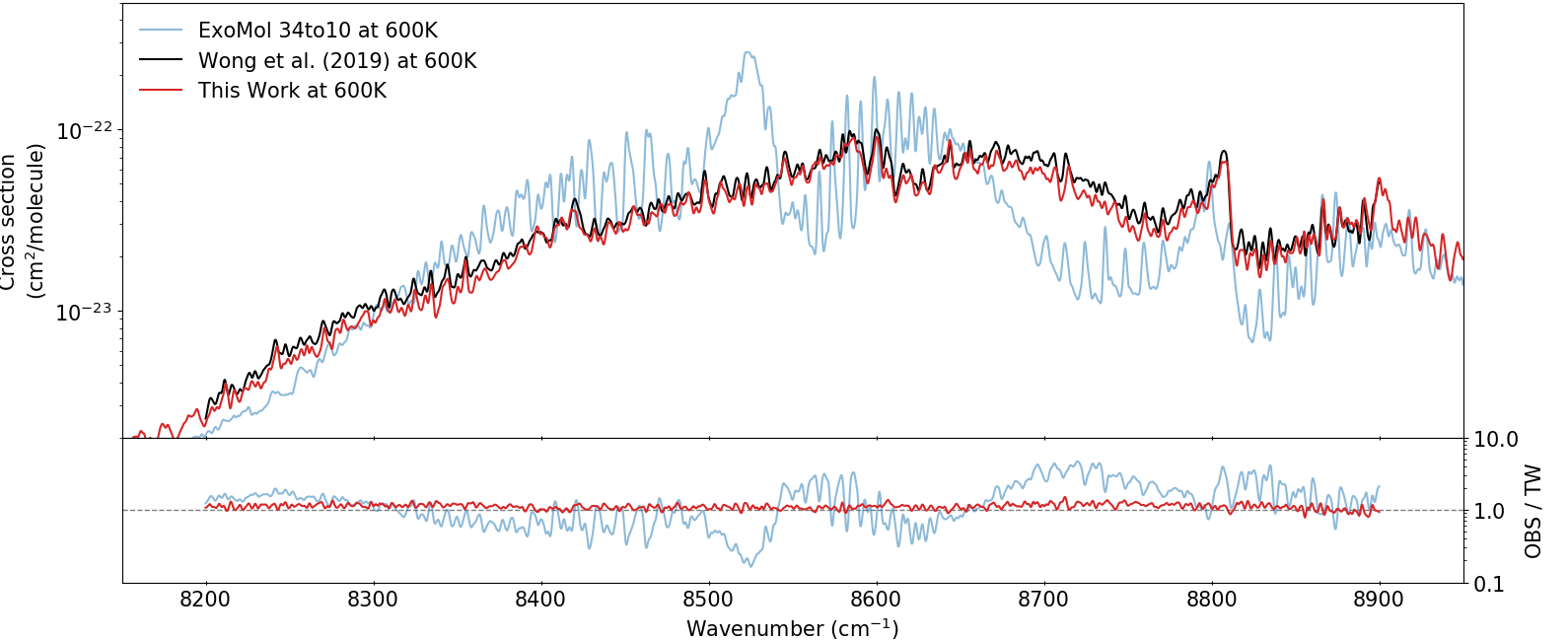}
\end{tabular}
\caption{Empirical absorption cross sections of CH$_{4}$ from \citet{2019ApJS..240....4W} at 600$\:$K, compared to calculations using this work and ExoMol 34to10 for the tetracontad (a), icosad (b), and triacontad (c) regions. In each case, the cross sections have been calculated for 100$\:$Torr of CH$_{4}$ and convolved to 2$\:$cm$^{-1}$. The lower panels in each plot display the residuals. \label{fig_full_conv_hot}}
\end{figure*}

\citet{2019ApJS..240....4W} have previously compared empirical absorption cross sections of the $P_{4}$, $P_{5}$ and $P_{6}$ regions of CH$_{4}$ to those calculated using RNT2017 up to 1000$\:$K. The largest spectral coverage provided by \citet{2019ApJS..240....4W} is at 600$\:$K, and comparisons to this work and ExoMol 34to10 are shown in Fig.~(\ref{fig_full_conv_hot}) for a convolved resolution of 2$\:$cm$^{-1}$. At this temperature and resolution, this work is able to reproduce almost all features seen in the $P_{4}$ and $P_{5}$ regions. For $P_{6}$, the overall shape of the observed band is reproduced, with a slight difference in intensity. The same comparisons were made for all temperatures available from \citet{2019ApJS..240....4W}, with similar results. For all bands, this work has significantly smaller residuals than ExoMol 34to10.

The window regions between the polyad bands become important for brown dwarf and exoplanet observations and are dominated by the effective lines (i.e., quasi-continuum) at high temperatures. It is extremely difficult to experimentally validate these regions due to the large pressures/path lengths required. However, comparisons with the \citet{2019ApJS..240....4W} measurements towards the edges of each region appears to imply that the intensities for these regions produced by this work (and therefore in RNT2017) are in better agreement than those from ExoMol 34to10.

\section{Discussion} \label{sec:discussion}

The fundamental objective of this work is to create a line list to be used for high-temperature applications, with HITRAN recommended at room-temperatures. However comparisons at 296$\:$K have been used to highlight the quality of the line list from this work. Figs.~(\ref{fig_hitran}) and (\ref{fig_hitran_conv}) demonstrate that this work is capable of accurately reproducing $^{12}$CH$_{4}$ absorption at 296$\:$K. This is to be partly expected, since this work contains substituted line parameters from the most accurate room-temperature line list available: HITRAN2016. For polyad bands up to tetradecad, calculated cross sections using this work compare extremely well with those using HITRAN2016. For the icosad and triacontad regions, the position accuracy reduces to around $\pm$1$\:$cm$^{-1}$, while intensities remain accurate to a few percent. For the tetracontad region and beyond, these accuracies are further reduced. As explained in Sect.~(\ref{subsec:lists:hitran}), the limited assignments and average/estimated values of $E^{\prime\prime}$ provided over these regions makes matching lines an insurmountable problem. For this reason, we strongly recommend using CH$_{4}$ from HITRAN2016 for room temperature applications.  It should be noted, that for calculations beyond 296$\:$K, the line intensities of HITRAN2016 lines with average/estimated values of $E^{\prime\prime}$ will lead to an increasing intensity error for increasing temperatures. This is an additional error due to the lack of hot bands and high rotational levels, but emphasizes the need to use an appropriate line list for the temperature range of interest.

For high temperature, this work has been compared to experimental observations up to 1173$\:$K, based on the work of \citet{2015ApJ...813...12H} and \citet{2019ApJS..240....4W}. For comparisons to the pentad and octad spectral regions, Figs.~(\ref{fig_pentad}) and (\ref{fig_octad}), this work is able to account for all features within a spectral resolution of $\pm$0.15$\:$cm$^{-1}$. Comparisons to \citet{2019ApJS..240....4W} experimental absorption cross sections at 600$\:$K, Fig.~(\ref{fig_full_conv_hot}), also show excellent agreement at a resolution of 2$\:$cm$^{-1}$. At higher temperatures, the coverage of the experimental cross sections provided by \citet{2019ApJS..240....4W} is reduced, which highlights the main difficultly of validating high-temperature line lists: a lack of observations. CH$_{4}$ also begins to dissociate towards higher temperatures, making it a challenge to record high-resolution spectra without also observing contamination species. Many more measurements are required throughout the infrared in order to further increase assignments (and thereby improve theoretical calculations), but also to further validate methane line lists. 

The comparisons presented in this work demonstrate that this line list contains the most accurate line parameters of CH$_{4}$ currently available for use at high temperatures. Indeed, a preliminary version of this work has already been applied to accurate remote-sensing measurements of CH$_{4}$ concentration in flames at 1000$\:$K \citep{2019arXiv191008116T}. Further validation is ongoing for high-temperature laboratory observations \citep{malarichn2019} and exoplanetary simulations \citep{roudier2019}. 

\begin{figure}[t!]
\centering
\includegraphics[scale=0.33, trim={0 0 0 0}, clip]{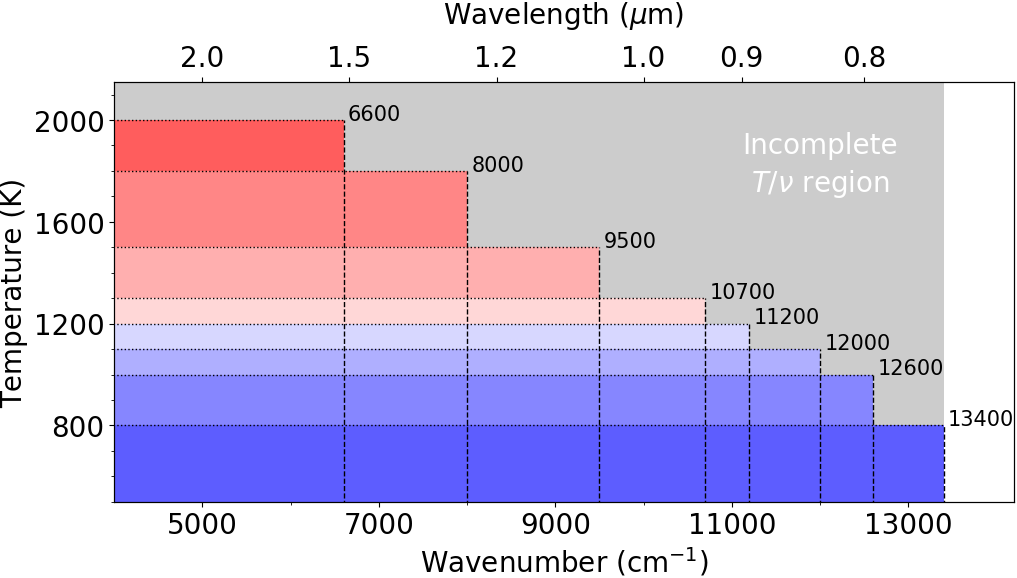}
\caption{A diagram for the completeness of the $^{12}$CH$_{4}$ HITEMP line list. Each colored region identifies the wavenumber range of completeness (up to the limits indicated) when used at high temperature. For $T<800\:$K, the HITEMP line list is complete between $\nu = 0$-13,400~cm$^{-1}$, whereas at 2000~K, it is complete between $\nu = 0$-6600~cm$^{-1}$. Use of the $^{12}$CH$_{4}$ HITEMP line list at temperatures/wavenumbers within the grey region will lead to errors as a consequence of incompleteness and is not recommended.  \label{fig_completeness}}
\end{figure}

The majority of lines contained in this work originate from RNT2017, and it is a testament to the accuracy of RNT2017 that this work agrees so well with observations up to 1173$\:$K. It should be stressed that the accuracy of line positions and intensities is limited to the spectral limits given in Tab.~(\ref{tab_ch4_strong_lines}) and has been illustrated in Fig.~(\ref{fig_completeness}) for the purposes of this work and therefore HITEMP. Using this work for spectral regions and temperatures that are outside of these bounds will lead to a reduction in accuracy, which increase with temperature. 

\begin{figure}[b!]
\centering
\begin{tabular}{c}
\includegraphics[scale=0.21, trim={0.2cm 3cm 0 0}, clip] {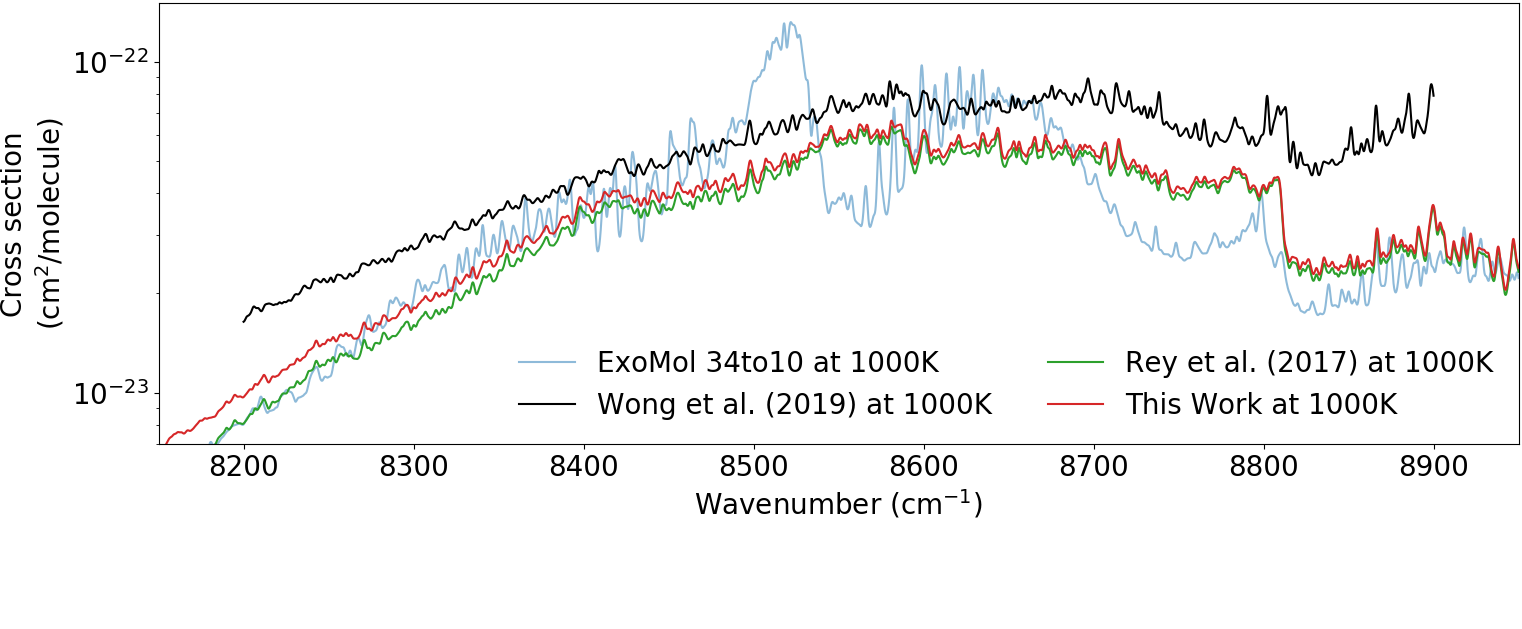} 
\end{tabular}
\caption{Empirical absorption cross sections of CH$_{4}$ from \citet{2019ApJS..240....4W} at 1000$\:$K, compared to calculations using this work, RNT2017 and ExoMol 34to10 for the triacontad region. In each case, the cross sections have been calculated for 100$\:$Torr of CH$_{4}$ and convolved to 2$\:$cm$^{-1}$.  \label{fig_full_conv_1000}}
\end{figure}

The effective lines included in this work can be considered a negligible source of error up to $\sim$1100$\:$K. For studies below this temperature, the contribution of the effective line absorption ($\Sigma S_{\textrm{\scriptsize{TWeff}}}(T)$ in Tab.~(\ref{tab_ch4_tw})) to the total absorption is seen to be small. Beyond 1100$\:$K, the contribution of the effective lines increases to almost 50\% at 2000$\:$K. Comparison with RNT2017 gives a maximum difference of 6\% at 1500$\:$K, with this difference primarily due to the effective line strengths. It is actually remarkable that the difference is only 6\% when considering that two effective lines are capable of reproducing the absorption between 300-2000$\:$K for a super-line grid point, each of which accounts for tens of thousands of lines. Fig.~(\ref{fig_full_conv_1000}) shows that when compared to \citet{2019ApJS..240....4W} at 1000~K, this increase in absorption from the effective lines actually brings this work closer to observations than the underlying RNT2017 line lists. The significant majority of lines in this region have not been assigned, meaning the uncertainty of line intensities can be quite large. This further highlights the need for additional observations to validate these intensities at higher temperatures.

In addition, the effective line uncertainties are a consequence of reanalyzing the RNT2017 super-line lists to remove the contribution of the strong lines at each temperature. On top of this, super-line grid points that are zero (or have had the intensity incorrectly removed) can introduce anomalies in the dual line fits described in Sect~(\ref{sec:combining}). The method used for this work is a non-ideal way to obtain the original RNT2017 line lists prior to compression. It can therefore be completely avoided by working with the original line lists. We again emphasize that original full line lists should be stored prior to super-line compression. We are already working towards an improved effective line approach, that can increase the accuracy of the effective line intensities over all temperatures by using an appropriate grid. 

Furthermore, \citet{2017A&A...605A..95Y} warn that a uniform super-line grid can introduce errors when calculating cross-sections for lower wavenumber regions: a consequence of the resolving power, $R$. They conclude that a constant resolving power grid gives the best results, where $R=\Delta\nu/\nu=100,000$. This provides $\sim$7 million grid points for the studied region. This work is restricted to the super-line lists provided by RNT2017, with a fixed grid spacing of 0.005~cm$^{-1}$. A constant resolving power grid will be considered for future updates to the CH$_{4}$ line list for HITEMP.

The effective lines are responsible for a continuum-like feature at all temperatures. At higher temperatures, the contribution of the effective lines increases and surpasses the absorption of the strong lines at higher wavenumbers, most noticeable for the window regions. Observations of transiting exoplanets are sensitive to these window regions (in addition to the band centers) as light passes through the limb of the planet's atmosphere. The contribution to these windows at high temperatures is caused by the continuum lines, and sensitive measurements of the CH$_{4}$ continuum-like absorption would be valuable for validation. Application of this work to simulations/retrievals of brown dwarfs and exoplanets will assist in the validation of these regions at the highest temperatures.

The completeness of RNT2017 is clearly described in \citet{2017ApJ...847..105R}, where efforts were made to keep each polyad complete up to the temperatures and spectral limits given in Tab.~(\ref{tab_ch4_strong_lines}). Our current work has been based on the RNT2017 line lists and therefore retains its completeness. The temperature-dependant wavenumber limits are also applicable to this work (and therefore the CH$_{4}$ HITEMP line list), which have been illustrated in Fig.~(\ref{fig_completeness}). For temperatures of spectral regions outside of these bounds, a scaling factor was recommended for the RNT2017 super-line lists to extrapolate the absorption and compensate for the lack of hot lines in these regions.  However, scaling of the effective lines is not recommended since discontinuities can appear at the wavenumber limits, as shown in Fig.~(\ref{fig_v_rnt2017}) at 12,600$\:$cm$^{-1}$. These discontinuities are a consequence of the effective lines being retrieved from fewer temperatures. For example, a super-line grid point $<$6600$\:$cm$^{-1}$ can be populated at all temperatures, whereas one $>$12,600$\:$cm$^{-1}$ is only populated up to 800~K, which reduces the quality of the effective line fits. For users who require the full spectroscopic detail of the complete RNT2017 line lists, we refer to the original work \citep{2017ApJ...847..105R} available from the TheoReTS data system \citep{2016JMoSp.327..138R}.

To create an accurate and complete line list of CH$_{4}$ requires the calculation of billions of transitions. This makes them difficult to use in their entirety, and RNT2017 (and ExoMol 34to10) have attempted to mitigate this effect by retaining a relatively small number of strong lines that account for the structure of polyad bands, while compressing the remaining calculated transitions into super-lines. However, these super-lines are temperature dependent and lack flexibility due to the loss of unique line information. For RNT2017, this significantly increases the total number of lines required to cover the full temperature range because a separate list is required at each temperature. In addition, this makes the lines lists difficult to use and are often converted into $k$-correlation tables to speed up atmospheric calculations \citep{2015ApJ...808..182G}. However, these approaches do not provide the flexibility and practicality of a single line list. 

The line list created for this work has been able to combine the accuracy of HITRAN2016 with the completeness of RNT2017 to form a single line list of CH$_{4}$ for high-temperature applications. This line list uses the familiar  HITRAN/HITEMP format, making it compatible with existing radiative-transfer software. Furthermore, the second generation of HAPI \citep{2016JQSRT.177...15K} is able to perform much faster radiative-transfer calculations by using ``just-in-time'' compiled codes while still using the line-by-line Voigt profile calculation with no interpolation. With this approach, the CH$_{4}$ line list consisting of $\sim$32~million lines shown in Fig.~(\ref{fig_total_list}) can be processed in approximately 450 seconds on a 12 core 2.6$\:$GHz CPU. These speed improvements make radiative-transfer calculations with the CH$_{4}$ HITEMP line list practical for all users.

Consequently, this work is currently the most accurate, and practical line list of CH$_{4}$ for high-temperature applications. This work has been submitted to \textit{Astrophysical Journal Supplement Series} and will made available via HITRAN\textit{online}, in the meantime the line list can be obtained by emailing the authors\footnote{robert.hargreaves@cfa.harvard.edu}. 

\section{Conclusion} \label{sec:conclusion}

This work has combined the separate $^{12}$CH$_{4}$ line lists of RNT2017 with HITRAN2016 to provide the most accurate line list of CH$_{4}$ for high-temperature applications. This work encompasses the 0-13,400~cm$^{-1}$ spectral region and is sufficiently complete to be used in line-by-line calculations up to 2000$\:$K. As a result, this work will be included as part of the HITEMP database and is currently available upon request.



To avoid incorrect conclusions during inter-comparisons with previous work, it is necessary to briefly summarize the difference in terminology and distinctions with respect to each other. The data compression strategy applied in the \textit{ab initio} born databases of RNT2017 \citep{2017ApJ...847..105R} and then in ExoMol 34to10 \citep{2017A&A...605A..95Y} consisted of a summation of weak line contributions within small wavenumber intervals, with the integrated features defined as super-lines. Typically, these super-lines are provided on a regular grid and include numerous transitions with a large variety of lower-state energy levels. Strong transitions were then provided as separate lists. The super-line approach permitted one to speed-up simulations of quasi-continuum absorption/emission cross-sections while the completeness was maintained. However, this strategy leads to the loss of information on individual lower-state energies. Thus a direct extrapolation of super-lines to other temperatures using standard conversion formulae is not possible. It was therefore necessary for RNT2017 and 34to10 to provide super-line lists for a range of temperatures.

In this present study, we adopt a different strategy for spectral data compression at high temperatures. To this end we have combined the most accurate theoretical calculations for CH$_{4}$ to date (i.e., RNT2017),  with HITRAN2016 parameters. Technically, this task is non-trivial due to separation between ``strong'' and ``weak'' transitions. In the TheoReTS database the strong line lists are temperature specific, accounting for the Boltzmann population of lower levels.  For the present HITEMP line list, we have combined all strong lines from RNT2017 at various temperatures into a single line list. That is to say, all transitions which could result in sufficiently sharp features at temperatures up to 2000$\:$K are included.  In addition, we have determined ``effective lines'' that can model the quasi-continuum features. These effective lines should not be confused with the previously described super-lines, even though they have been determined from the super-line lists. The main difference is that we have attributed an effective lower-state energy to each of the  effective lines (i.e., it does not correspond to true quantum state or individual ro-vibrational transition), which enables them to be converted to any temperature using the standard formulae. The primary advantage of this strategy is the effective lines can be included alongside the strong lines to form a single, practical line list of CH$_{4}$. This strategy has enabled the production of a line list containing $\sim$32 million lines, provided in the HITRAN/HITEMP format and that is compatible with existing radiative-transfer software. This work is able to reproduce the RNT2017 intensities up to 2000$\:$K, but it must be stressed that using this work outside of the recommended temperature/wavenumber ranges may lead to issues of completeness. The effective lines have to be used with caution as they are convenient for radiative-transfer simulations but should not be used for calculating energy levels as they do not represent transitions between real quantum states.   

The HITEMP line list of CH$_{4}$ produced for this work has been validated against available high-temperature measurements of CH$_{4}$ polyads up to 9000$\:$cm$^{-1}$ ($>$1.11$\:\mu$m), the triacontad region. Comparisons to alternative CH$_{4}$ line lists demonstrate that this work is the most accurate at reproducing observations at high temperatures.  

The majority of the lines in our CH$_{4}$ line list are due to the principal isotopologue of methane; however the line list has been supplemented with lines of $^{13}$CH$_{4}$, $^{12}$CH$_{3}$D and $^{13}$CH$_{3}$D taken from HITRAN2016.  In the future, it will be productive to include improved high-temperature line lists for these isotopologues, although they are not expected to be dominant absorption features in  CH$_{4}$ spectra.

\section{SUPPLEMENTARY DATA} \label{sec:supplemental}
The line list will be made available via HITRAN\textit{online} but is currently available upon request by emailing the authors.

\acknowledgments

We would like to acknowledge J. A. Karns for contributions made towards Python data input routines. Update of the HITRAN and HITEMP databases was supported through the NASA Aura and PDART grants NNX17AI78G and NNX16AG51G, respectively. Support from the French ANR e-PYTHEAS project and from the D.~Mendeleev program of Tomsk State University are also acknowledged.


\bibliography{CH4_HITEMP}{}
\bibliographystyle{aasjournal}

\end{document}